\documentclass{aa} 

\usepackage{graphicx}
\usepackage{ulem}
\usepackage[T1]{fontenc}
\usepackage{tabularx}
\usepackage{txfonts}
\usepackage{gensymb}

\usepackage[colorlinks=true, allcolors=blue]{hyperref}
\hypersetup{
     colorlinks   = true,
     citecolor    = blue,
     urlcolor = blue
}

\begin{document} 

    \title{The Northern Cross Fast Radio Burst project}

   \subtitle{V. Search for transient radio emission from Galactic magnetars}

%

   \author{A.~Geminardi\inst{1,2,3} 
          \and
          P.~Esposito\inst{1,4} 
          \and
          G.~Bernardi\inst{5,6,7}
          \and
          M.~Pilia\inst{3}
          \and
          D.~Pelliciari\inst{5}
          \and
          G.~Naldi\inst{5}
          \and
          D.~Dallacasa\inst{5,8} 
          \and
          R.~Turolla\inst{9,10} 
          \and
          L.~Stella\inst{11} 
          \and
          F.~Perini\inst{5}
          \and
          F.~Verrecchia\inst{11,12}
          \and
          C.~Casentini\inst{13,14}
          \and
          M.~Trudu\inst{3}
          \and
          R.~Lulli\inst{5}
          \and
          A.~Maccaferri\inst{5}
          \and
          A.~Magro\inst{15}
          \and
          A.~Mattana\inst{5}
          \and          
          G.~Bianchi\inst{5}
          \and
          G.~Pupillo\inst{5}
          \and
          C.~Bortolotti\inst{5} 
          \and
          M.~Tavani\inst{13,16} 
          \and
          M.~Roma\inst{5}
          \and
          M.~Schiaffino\inst{5}
          \and G.~Setti\inst{5,8}
}

   \institute{Scuola Universitaria Superiore IUSS Pavia, Palazzo del Broletto, piazza della Vittoria 15, I-27100 Pavia, Italy\\
              \email{andrea.geminardi@iusspavia.it}
        \and
            Dipartimento di Fisica, Universit\`a di Trento, via Sommarive 14, I-38123 Povo (TN), Italy
        \and
             INAF--Osservatorio Astronomico di Cagliari, via della Scienza 5, I-09047, Selargius (CA), Italy
        \and
            INAF--Istituto di Astrofisica Spaziale e Fisica Cosmica di Milano, via Corti 12, I-20133 Milano, Italy
        \and
            INAF--Istituto di Radio Astronomia (IRA), via Piero Gobetti 101, I-40129 Bologna, Italy
        \and
             South African Radio Astronomy Observatory, Black River Park, 2 Fir Street, Observatory, Cape Town, 7925, South Africa
        \and
             Department of Physics and Electronics, Rhodes University, PO Box 94, Makhanda, 6140, South Africa
        \and
            Dipartimento di Fisica e Astronomia, Universit\`a di Bologna, via Gobetti 93/2, I-40129 Bologna, Italy
        \and
            Dipartimento di Fisica e Astronomia, Universit\`a di Padova, via Marzolo 8, I-35131 Padova, Italy
        \and 
            Mullard Space Science Laboratory, University College London, Holmbury St Mary, Dorking, Surrey RH5 6NT, UK
        \and
            INAF--Osservatorio Astronomico di Roma, via Frascati 33, I-00078 Monteporzio Catone, Italy
        \and
            ASI--Space Science Data Center, via del Politecnico snc, I-00133 Roma (RM), Italy;
        \and
            INAF--IAPS, via del Fosso del Cavaliere 100, I-00133 Roma (RM), Italy;
        \and
            INFN Tor Vergata, via della Ricerca Scientifica, I-00133 Roma (RM), Italy
        \and
            Institute of Space Sciences and Astronomy (ISSA), University of Malta, Msida, MSD 2080, Malta
        \and
            Dipartimento di Fisica, Universit\`a degli Studi di Roma "Tor Vergata", via della Ricerca Scientifica 1, I-00133 Roma (RM), Italy}
            
   \date{XXX-XXX-XXX}

 
  \abstract
   {The radio emission from magnetars is poorly understood and poorly characterized observationally, in particular for what concerns single pulses and sporadic events. 
   The interest in it was boosted by the detection in 2020 of an extremely bright ms radio signal from the Galactic magnetar designated Soft Gamma Repeater (SGR) SGR\,J1935+2154, which occurred almost simultaneously with a typical magnetar short burst of X-rays.
   As of now, this event remains the Galactic radio pulse that is the most reminiscent of fast radio bursts (FRBs) and the only one with a sound association with a known progenitor.}
   {We aim to constrain the rate of impulsive radio events from magnetars, by means of an intensive monitoring using a high-sensitivity radio telescope.}
   {We performed a long-term campaign on seven Galactic magnetars (plus one candidate) using the Northern Cross transit radio telescope (in Medicina, Italy) searching for short timescales and dispersed radio pulses.}
   {We obtained no detections in $\sim$560 hours of observation, setting an upper limit at 95\% confidence level of $<$52 yr$^{-1}$ on the rate of events with energy $\gtrsim$$10^{28}$\,erg, which is consistent with limits in literature.
   Furthermore, under some assumptions on the magnetars properties and energetic behavior, we found that our upper limits point towards the fact that the entire population of FRBs observed cannot be explained by radio bursts emitted by magnetars.
   }
   {}

   
   \titlerunning{V}

   \maketitle
\nolinenumbers
\section{Introduction}\label{sec:Intro}
Magnetars are a class of neutron stars characterized by their high magnetic field strength ($10^{14}$--$10^{15}$ G) and by their peculiar emission behavior \citep{Mereghetti_magnetars_2008,Turolla_magnetars_2015,Kaspi_magnetars,Esposito_magnetars_2021}.
These compact objects can exhibit persistent and transient X-ray emission and, in some cases, also optical and infrared counterparts (\cite{0501_counter_Chrimes} and references therein). 
The signature feature of magnetars is their bursting activity. In addition, their emission and timing properties can undergo enormous changes over different timescales and these sharp variations are unpredictable \citep{Rea_esposito_outbursts}.
The longest magnetar events are the outburst phases, where the X-ray luminosity grows by a few orders of magnitude with respect to the quiescent phases and can last from weeks to years \citep{Coti_Zelati_outbursts}.
On shorter timescales, magnetars exhibit explosive events such as short bursts ($\lesssim$1\,s and luminosity $L\approx$$10^{36}$--$10^{43}$\,erg\,s$^{-1}$), which are often clustered in activity windows that can last for days \citep{Gogus_burts_2001,Israel_bursts_2008,Esposito_2008_1627_outburst}, and the rarest and extremely energetic giant flares. In the latter, the initial ms gamma-ray flash  ($L\approx10^{45}$--$10^{47}$\,erg\,s$^{-1}$) is followed by a fading X/$\gamma$-ray afterglow modulated at the neutron star spin period that can last for minutes and with characteristic total energy of $\approx$$10^{44}$\,erg \citep{Mazets_1979_Giant_flare,SGR1900_Hurley_A,Palmer_MGF_SGR1806-20}.
So far, the number of Galactic sources identified as magnetars is $\sim$30 \citep{Olausen_McGill_Cat}.\footnote{See \href{https://www.physics.mcgill.ca/~pulsar/magnetar/main.html}{\texttt{www.physics.mcgill.ca/pulsar/magnetar/main.html}}.}
However, the number of these sources in our Galaxy is expected to be much higher (\citealt{Mag_pop_Muno,Mag_pop_Gullon}; Ronchi et al., in preparation).
At radio wavelengths, ephemeral pulsations have been detected only from five magnetars during outburst phases \citep[][and references therein]{Rea_J1745_radio,Levin_J1622_radio,Camilo_XTEJ1810_radio,Camilo_1E1547_radio,Esposito_J1818_radio}.\footnote{Possibly six, including SGR\,J1935+2154, which was detected only with FAST in a small fraction of their monitoring time \citep{1935_pulsed_emission_FAST}.}
The energy involved in their radio emission is comparable or higher with respect to standard rotation-powered pulsars, while some clear different features distinguish the bursting behaviors of these two classes \citep{Kramer_magnetar_emission}.
Radio pulses coming from magnetars present a harder spectrum ($\propto \nu^{-0.5}$), in particular, the short timescale and unpredictable variations of the shapes of the pulses do not permit to identify a characteristic pulsing profile.
Owing to their extreme properties and bursting behavior, magnetars are often invoked in various astrophysical phenomena, such as gamma-ray bursts or super-luminous supernovae engines and ultra-luminous X-ray sources \citep{Dallosso_magnetars}.
In recent years, one of the most intriguing proposed magnetar association is that with the so-called fast radio bursts \citep{Popov_magnetars_FRB,Bailes_FRB_review,Zhang_FRB_review}.

FRBs are powerful millisecond bursts of celestial origin detected only at radio frequencies.
Their high spectral luminosity, typically greater than $10^{30}$\,erg s$^{-1}$ Hz$^{-1}$, and extreme brightness temperature ($\gtrsim$$10^{30}$\,K, with some of them exceeding $10^{36}$\,K) suggest coherent emission mechanisms in compact sources \citep{Nimmo_2022_frb}.
Among the great variety of models proposed to explain these intriguing signals, those involving the magnetosphere of magnetars or shocks beyond the magnetosphere are among the most popular and promising \citep{Platts_FRB_models}.
Since their discovery in 2007 \citep{Lorimer2007}, more than $800$ FRB sources have been found \citep{Blinkverse_cat} and current estimations suggest a high all-sky rate of FRBs $ \sim$$500$ per day above a $5\ \mathrm{Jy\,ms}$ threshold at $600$ MHz \citep{CHIME_CAT1_corrected}. 
FRBs show high flux densities (few Jy, reaching $\sim\mathrm{kJy}$ level in few cases \citep{Baseband_chime_catalog}), and their characteristic high dispersion measure (DM), which far exceeds the predicted Galactic contribution in the directions of arrival of the FRBs, points to an extragalactic origin. Some associations with galaxies, with the most distant detected so far at redshift $z \sim 1$ \citep{Ryder_2023_Z1}, suggest that FRBs can come from great distances and could be exploited in cosmological studies \citep{FRB_cosmo}.

The main subdivision in the FRB population is between sources detected only once (one-off FRBs) and those that showed recurrent activity (FRB repeaters), which are the $\sim$$8 \%$ of the known population \citep{Blinkverse_cat}. \footnote{However, \cite{CHIME_repeaters_perc} suggest a repeaters detection rate of $\sim$3\%.}
Our current knowledge and understanding of FRBs is not sufficient to assess whether repeating sources represent a standalone population with respect to one-off FRBs and not even whether they are produced by the same sources or mechanisms \citep{Pleunis_2021_frbcat}.
Today, despite many efforts, there are no counterparts associated with any known extragalactic FRB.

The suggested link between magnetars and FRBs was strengthened by the peculiar radio activity that has been observed from SGR\,J1935+2154, which was an ``ordinary'' magnetar until that point.
In 2020, this source emitted several bright, millisecond radio bursts \citep{SGR1935_CHIME,SGR1935_Bochenek,SGR1935_Kirsten,SGR1935_Rehan}. One of these was extremely bright \citep{SGR1935_CHIME}, lasted for $\sim$1\,ms, reached a fluence of 480\,kJy\,ms at 400 -- 800 MHz and a MJy\,ms fluence at 1.4 GHz.
Up to date, this is the Galactic radio signal that is most reminiscent of FRBs, although the inferred energy released was few order of magnitude lower than the typical FRBs energy.

In this work, we aim to constrain the radio-bursting activity of magnetar-like objects.
We focused our search on millisecond-long transients coming from Galactic magnetars with intense monitoring at high sensitivity.
Our campaign is sensitive to both FRB-like events and giant pulses coming from the observed magnetars \citep{Crab_giant_pulses,J1810_Giant_pulses,Camilo_XTEJ1810_radio,Israel_1547_2021}.
Furthermore, under some assumptions, we exploited our observations to investigate the magnetar energy spectrum and their high-energy behavior.

This paper is summarized as follows. Section \ref{sec:Targets} describes the targets observed in this work. In Section \ref{sec:Obervations}, we outline how the data are stored and analyzed. Section \ref{sec:methods} contains the model used to obtain the results that are listed and discussed in Section \ref{sec:results}. Finally, in Section \ref{sec:conclusions} we discuss the results and summarize our work.

\section{Targets} \label{sec:Targets}

\begin{table*}
    \caption{Localization and predicted Galactic DM contributions for the magnetar sample.}
    \resizebox{\textwidth}{!}{
    \renewcommand{\arraystretch}{1.5}
    \begin{tabular}{l c c c c c c c c c}
        \hline
        \hline
        Name & Gal L & Gal B & D & n$_{\rm{H}}$ & DM$_{\rm{NE2001}}$& DM$_{\rm{YMW16}}$  & DM$_{\rm{nH}}$ & DM$_{\rm{radio}}$  & Ref. \\
        & (deg) & (deg) & (kpc) & ($10^{22}$ cm$^{-2}$) & (pc cm$^{-3}$) & (pc cm$^{-3}$) & (pc cm$^{-3}$) & (pc cm$^{-3}$) & \\
        \hline
        3XMM\,J185246.6+003317 & 33.5785548 & -0.0450683 & $7.1 \pm 2.1$ & $1.32 \pm 0.05$ & $441^{+275}_{-274}$ & $1020^{+590}_{-633}$ & $440 \pm 16$ & No & 1 \\

        SGR\,1900+14 & 43.0034808 & 0.8520327 & $12.5 \pm 1.7$ & $2.12 \pm 0.08$ & $609^{+55}_{-80}$ & $526^{+70}_{-105}$ & $707 \pm 26$ & No & 2,3 \\

        SGR\,J1935+2154 & 57.2467449 & 0.8189765 & $9.0 \pm 2.5$ & $1.6 \pm 0.2$ & $293^{+100}_{-114}$ & $320^{+90}_{-119}$ & $533 \pm 67$ & $332.7206 \pm 0.0009$ & 4,5,6 \\

        SGR\,2013+34 & 72.3203222 & -0.1010242 & $8.8 \pm 2.4$ & -- & $279^{+77}_{-110}$ & $302^{+70}_{-133}$ & -- & No & 7 \\

        1E\,2259+586 & 109.0873535 & -0.9957528 & $3.2 \pm 0.2$ & $1.21 \pm 0.04$ & $99^{+11}_{-11}$ & $154^{+13}_{-18}$ & $403 \pm 13$ & No & 8,9 \\

        4U\,0142+61 & 129.3839879 & -0.4307465 & $3.6 \pm 0.4$ & $1.00 \pm 0.01$ & $96^{+10}_{-11}$ & $161^{+5}_{-6}$ & $333 \pm 3$ & No & 10,11 \\

        SGR\,0418+5729 & 147.9790422 & 5.1191370 & $2.0 \pm 0.6$ & $0.115 \pm 0.006$ & $63^{+26}_{-25}$ & $93^{+32}_{-55}$ & $38 \pm 2$ & No & 12,13 \\

        SGR\,0501+4516 & 161.5466873 & 1.9489249 & $2.0 \pm 0.6$ & $0.88 \pm 0.01$ & $71^{+27}_{-28}$ & $109^{+27}_{-61}$ & $293 \pm 3 $ & No & 14,15 \\
        \hline
    \end{tabular}}
    \textbf{Notes.} The second and third columns list the Galactic latitude and longitude of the targets, respectively.
    The fourth column contains the distances adopted for the magnetars; the distances of 3XMM\,J185246.6+003317, SGR\,2013+34, SGR\,0418+5729 and SGR\,0501+4516 are strongly discussed in the literature, therefore we choose conservative uncertainty intervals of $\pm 30 \% $ around the most credible values.
    The fifth column represents the hydrogen column density derived from high-energy observations.
    The sixth and seventh columns show the inferred DM contributions, assuming our values for the distances, using the electron density distribution models \cite{Pyne2001} and \cite{YMW16}, respectively.
    The eighth column contains the DM contribution obtained with the best-fit relation between n$_{\rm{H}}$ and DM in \cite{HE_NH_DM_2013}. Finally, the last column shows the DM obtained with radio observations; this value is available only for SGR\,J1935+2154, as the other magnetars of our sample have never been detected at radio wavelengths.\\
    \textbf{References.} [1] \cite{3XMMJ18+00_REA}; [2] \cite{SGR1900+14_MEREGHETTI}; [3] \cite{SGR1900+14_DAVIES}; [4] \cite{SGR1935_Zhong}; [5] \cite{SGR1935+2154_ISRAEL}; \cite{SGR1935_CHIME}; [7] \cite{SGR2013+34_SAKAMOTO}; [8] \cite{1E2259+586_KOTHES}; [9] \cite{1E2259+586_PIZZOCARO}; [10] \cite{4U0142+61_DURANT}; [11] \cite{4U0142_REA}; [12] \cite{SGR0418_VDH_2010}; [13] \cite{SGR0418_REA_2013}; [14] \cite{SGR0501_Lin}; [15] \cite{SGR0501_Camero}.
    \label{tab:Magnetars_info}
\end{table*}

In this section, we summarize the properties of the sample of magnetars that we observed. We monitored all the currently known Galactic magnetars that are visible from the Northern Cross (NC) site in Medicina, near Bologna (Italy) \citep{Olausen_McGill_Cat}. Most of the targets are located in the direction of the Galactic plane; Table \ref{tab:Magnetars_info} shows the coordinates and estimated DMs for each source.
We perform DM estimations using the models of \cite{Pyne2001} and \cite{YMW16} based on the electron density distribution in our Galaxy and also using the best fit of the relation between DM and the observed hydrogen column density of \cite{HE_NH_DM_2013}.

\subsection{3XMM\,J185246.6+003317}
3XMM\,J185246.6+003317 was discovered and identified as a magnetar in 2014 \citep{3XMMJ18_Zhou}. From the re-analysis of \textit{XMM-Newton} data, this source showed an outburst phase between 2008 and 2009. The neutron star 
rotates with a period of $P\sim 11.56 $\,s and the upper limit on the spin period derivative is $\dot{P} < 1.4 \times 10^{-13}$\,s\,s$^{-1}$, implying a dipolar surface magnetic field of $B_{\rm{dip}} < 4.1 \times 10^{13}$ G \citep{3XMMJ18+00_REA}.
The latter are typical parameters of a low magnetic field (low-B), aged magnetar.
No radio signals have ever been detected from this source.
3XMM\,J185246.6+003317 is located at a distance of $1'$ from the supernova remnant Kes 79 and the absorption properties are similar between the two objects. Therefore, as in \cite{3XMMJ18_Zhou}, we assume $ \sim$$7.1 $ kpc as the distance for both Kes 79 and magnetar. The sky coordinates used for the observations of this target are R.A. = $18^{\rm h}52^{\rm m} 46.67^{\rm s}$, Dec $=+00^\circ33'17.8''$ (J2000).

\subsection{SGR\,1900+14}
SGR\,1900+14 was discovered in 1979 \citep{SGR1900_Mazets}, it is a persistent X-ray source, it has spin period $P\sim 5.20 $ s and period derivative $\dot{P} \sim 9.2 \times 10^{-11}$ s s$^{-1}$ \citep{SGR1900+14_MEREGHETTI}; the inferred surface magnetic field is $B_{\rm{dip}} \sim 8 \times 10^{14}$ G \citep{SGR1900_Kouveliotou_1999}.
SGR\,1900+14 emitted a giant flare in 1998 \citep{SGR1900_Hurley_A} reaching a total energy of $\sim 10^{44}$ erg and in the following years several high-energy bursts have been detected from this target. Very Large Array (VLA) observations associated a fading radio source with this object just after the giant flare \citep{SGR1900_Frail}. Similarly to what was observed for SGR\,1806-20 after its 2004 giant flare \citep{Cameron_1806_prs}, the radio source was tentatively attributed to an outflow of relativistic particles. 
The distance from Earth of this magnetar is not well established. \cite{SGR1900_Kouveliotou_1999} and \cite{SGR1900_Hurley} estimate a distance between 5 and 7 kpc, while others, e.g. \cite{SGR1900+14_MEREGHETTI} and \cite{SGR1900+14_DAVIES}, locate the magnetar between 12.5 and 15 kpc. Accordingly with the kinematic distance estimated in \cite{SGR1900+14_DAVIES}, we assume a distance of $12.0 \pm 1.7$ kpc and the sky coordinates R.A. = $19^{\rm h}06^{\rm m} 53.7^{\rm s}$, Dec $=+09^\circ20'47.0''$ (J2000).

\subsection{SGR\,J1935+2154}
SGR\,J1935+2154 was identified as a magnetar in 2015 \citep{SGR1935+2154_ISRAEL}, without a detected radio counterpart.
The pulsation period is $P\sim 3.24$ s and the spin-down rate $\dot{P} \sim 1.43 \times 10^{-11}$ s s$^{-1}$, resulting in an estimated surface dipolar magnetic field of $B_{\rm{dip}} \sim 2.2 \times 10^{14}$ G \citep{SGR1935+2154_ISRAEL}.
This magnetar is of particular interest because it exhibits subsequent outburst phases since its discovery \citep[e.g.][]{Borghese_1935,Ibrahim_1935} and because of its transient radio activity with several detected radio bursts.
In particular, in 2020, SGR\,J1935+2154 emitted an FRB-like event \citep{SGR1935_CHIME} and a few others isolated radio bursts \citep{SGR1935_Kirsten,SGR1935_Rehan}, showing a dispersion measure of $\rm{DM}_{\rm{J1935}} \sim 332$ pc cm$^{-3}$. The FRB-like event was detected during an intense X-ray activity phase and it arrived in advance of $6.5 \pm 1.0$ ms with respect to a bright X-ray burst \citep{SGR1935_Mereghetti,1935_Tavani,SGR1935_Bochenek}.
The estimate of the distance of SGR\,J1935+2154 is still an open question. The supernova remnant G57.2+0.8 has been proposed as the origin of this magnetar, with inferred distances spanning from $4.4$ to $12.5$ kpc \citep{SGR1935_Kothes_dist,SGR1935_Ranasinghe_dist}. Exploiting the observational constraints and the modeling of the DM distribution in the direction of SGR\,J1935+2154, as in \cite{SGR1935_Zhong}, we assume a distance of $9.0 \pm 2.5$ kpc. The sky coordinates used for the observations of this target are R.A. = $19^{\rm h}34^{\rm m} 59.5978^{\rm s}$, Dec $=+21^\circ53'47.7864''$ (J2000).

\subsection{SGR\,2013+34}
SGR\,2013+34 is a candidate magnetar associated to the detection of an X-ray signal (GRB 050925) in 2005 \citep{SGR2013_SWIFT_CAT} of $\sim 90 $ ms of duration coming from a direction in the Galactic plane.
The inspection of the BATSE Gamma-Ray Burst Catalog\footnote{See \href{http://www.batse.msfc.nasa.gov/batse/grb/catalog/current/}{\texttt{www.batse.msfc.nasa.gov/batse/grb/catalog/current/}}.} produced 3 bursts possibly consistent with the direction of GRB 050925 \citep{SGR2013+34_SAKAMOTO}.\footnote{The typical localization accuracy is $\sim$2$^{\degree}$ (68$\%$ containment radius), see \hyperlink{https://gammaray.nsstc.nasa.gov/batse/grb/rbr/}{\texttt{https://gammaray.nsstc.nasa.gov/batse/grb/rbr/}}.}
These few clues are not enough to establish if this target is a magnetar or not, we observed this source for completeness and consistency with the \cite{Olausen_McGill_Cat} catalog.
In the direction of arrival of GRB 050925 there are few $\rm{H}_{\rm{II}}$ regions, assuming that the source is associated to one of them, the inferred distance is $8.8$ kpc \citep{SGR2013+34_SAKAMOTO}. The sky localization of GRB 050925 is R.A. = $20^{\rm h}13^{\rm m} 56.9^{\rm s}$, Dec $=+34^\circ19'48.0''$ (J2000).

\subsection{1E\,2259+586}
1E\,2259+586 was discovered in 1981 owing to its bright X-ray pulsating behavior \citep{1E2259_Fahlman} which exceeded the energy that a common rotation-powered neutron star is able to emit.
In 2002, 1E\,2259+586 entered an outburst phase with its persistent X-ray flux increased by a factor $\sim 10$, showing also a bursting activity and one glitch (a timing anomaly frequently associated with magnetars; \citealt{Woods_2004_1E2259_outburst}). Furthermore, another outburst was detected in 2012 \citep{1E2259_Foley_outburst} showing an increased soft X-ray flux, strongly suggesting that the two classes of anomalous X-ray pulsars (AXPs) and soft gamma repeaters (SGRs) were actually parts of the same population.
This magnetar has a period of pulsation of $P\sim 6.98$ s, spin derivative of $\dot{P} \sim 4.84 \times 10^{-13}$ s s$^{-1}$ and the inferred surface magnetic field strength is $B_{\rm{dip}} \sim 5.9 \times 10^{13}$ G \citep{1E_2259_Dib_2014}.
1E\,2259+586 is associated to the supernova remnant CTB109 in the Galactic Perseus spiral arm located at a distance of $3.2 \pm 0.2$ kpc \citep{1E2259+586_KOTHES} and its sky coordinates are R.A. = $23^{\rm h}01^{\rm m} 08.295^{\rm s}$, Dec $=+58^\circ52'44.45''$ (J2000).

\subsection{4U\,0142+61}
4U\,0142+61 is one of the brightest persistent X-ray magnetars ever detected \citep{4U0142_ISRAEL}. It showed also optical \citep{4U0142_Hulleman_2000} and infrared counterparts \citep{4U0142_Hulleman_2004}. 
This magnetar has a pulse period of $P\sim 8.69$ s with spin derivative $\dot{P} \sim 2 \times 10^{-12}$ s s$^{-1}$ that imply a surface dipolar magnetic field of $B_{\rm{dip}} \sim 6 \times 10^{13}$ G \citep{1E_2259_Dib_2014}. A possible glitch-like event \citep{4U0142_MORII} was reported for 4U\,0142+61 and it also showed periods of high bursting activity.
This magnetar is located at a distance of $3.6 \pm 0.4$ kpc \citep{4U0142+61_DURANT,Tendulkar_2013_4U0142_1E2259} at sky coordinates R.A. = $01^{\rm h}46^{\rm m} 22.407^{\rm s}$, Dec $=+61^\circ45'03.19''$ (J2000).

\subsection{SGR\,0418+5729}
SGR\,0418+5729 was discovered in 2009 during an outburst phase \citep{SGR0418_VDH_2010}.
This is a low-B magnetar, its inferred surface dipolar magnetic field is $B_{\rm{dip}} \sim 6 \times 10^{12}$ G due to a period of pulsation $P\sim 9.08$ s and spin derivative of $\dot{P} \sim 4 \times 10^{-15}$ s s$^{-1}$ \citep{Esposito_0418,SGR0418_REA_2010,SGR0418_REA_2013}.
The well studied spin-down and the low magnetic field of this source provided the first confirmations that the dipolar component of the magnetic field of magnetars is not enough to explain their behavior, suggesting a twisted and strong multi-polar structure.
No radio and other counterparts were detected for this target.
SGR\,0418+5729 is arguably located at a distance of $\sim 2$ kpc in the Perseus arm \citep{Perseus_Xu} at sky coordinates R.A. = $04^{\rm h}18^{\rm m} 33.867^{\rm s}$, Dec $=+57^\circ32'22.91''$ (J2000) \citep{SGR0418_VDH_2010}.

\subsection{SGR\,0501+4516}
SGR\,0501+4516 is a magnetar discovered by the \textit{Neil Gehrels Swift Observatory} satellite in 2008 \citep{SGR0501_Barthelmy,SGR0501_Rea} during an intense bursting activity related to an outburst phase.
It has a spin period of $P\sim 5.76$ s, spin derivative of $\dot{P} \sim 5.94 \times 10^{-12}$ s s$^{-1}$ and a derived surface dipolar magnetic field of $B_{\rm{dip}} \sim 1.9 \times 10^{14}$ G \citep{SGR0501_Camero}.
Both an optical/infrared counterpart and optical pulsation \citep{SGR0501_Dhillon} have been detected for this magnetar.
The estimate of the distance is the same of SGR\,0418+5729, both the sources are located in the Perseus arm \citep{Perseus_Xu} with the most reliable distance of $2$ kpc \citep{SGR0501_Lin}.
The sky localization is R.A. = $05^{\rm h}01^{\rm m} 06.76^{\rm s}$, Dec $=+45^\circ16'33.92''$ (J2000).

\section{Observations} \label{sec:Obervations}
The NC is a T-shaped transit radio telescope situated in Medicina (BO), Italy \citep{NC_Project_1}. It is constituted by two perpendicular arms orientated along the East-West and North-South directions.
The East-West arm has a single cylindric-parabolic antenna of size $564 \times 35$ m with a total of 1488 dipoles.
The North-South arm is an array of 64 receiving cylinders, each of size $7.5 \times 23.5$ m, corresponding to a total collecting area of $\sim$11280 m$^2$. A great part of the telescope is under refurbishment and the observations presented in this work have been performed using the currently available receivers (i.e. 16 cylinders of the North-South arm), as in \cite{NC_Project_4}. The instrument produces down-sampled data with a time resolution of $\tau_{\rm{s}} = 138 \; \mu$s, over an operative bandwidth of $\sim 14.8$ MHz, divided into 1024 frequency channels and centered at 408 MHz.

\begin{figure}[h]
    \centering
    \includegraphics[width=\linewidth]{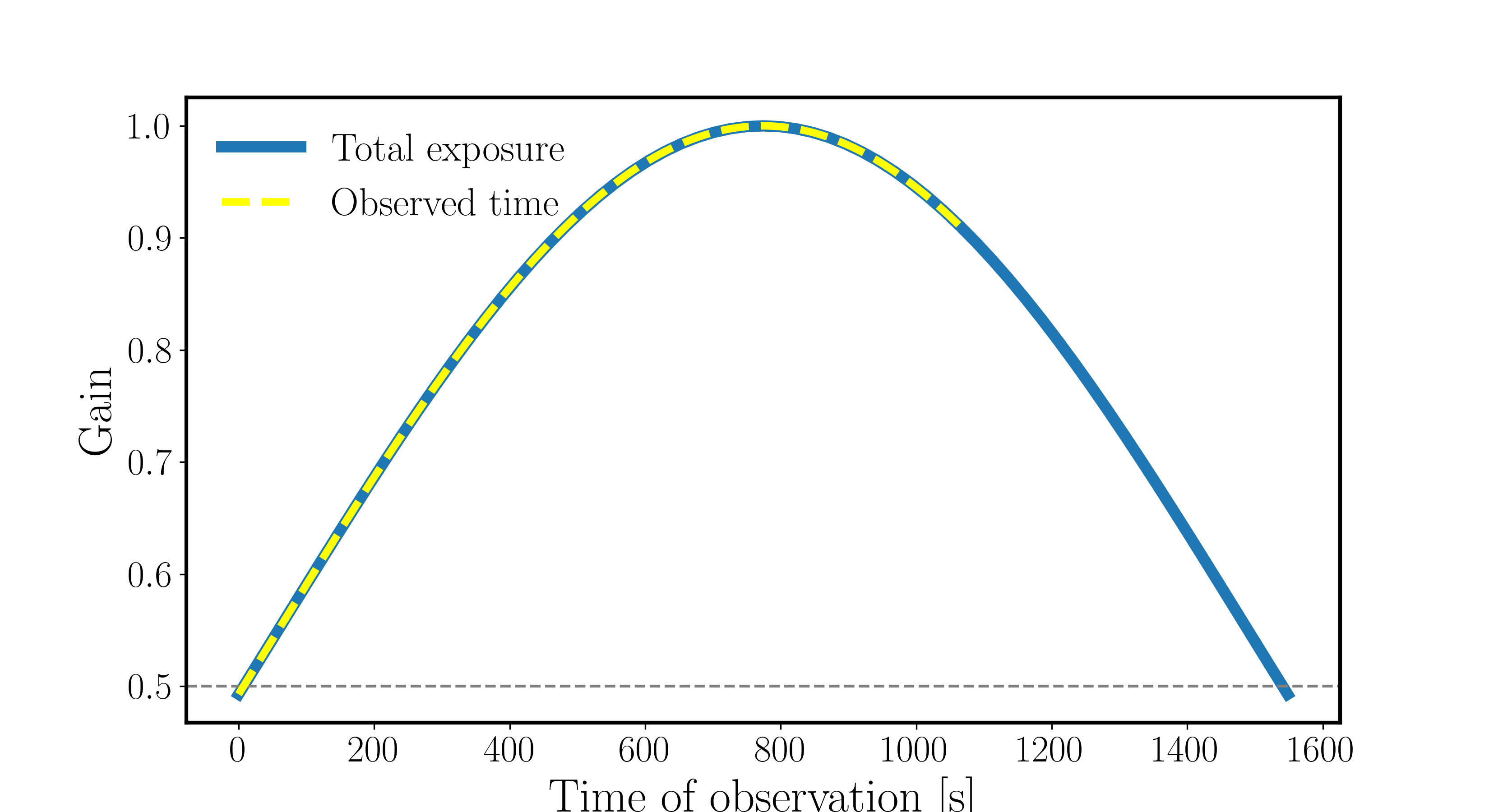}
    \caption{Gain profile of a NC observation on 3XMM\,J185246.6+003317, as an example. The target has a total exposure of 1548 s every day. The latter is defined as the amount of time per day during which the NC can observe the target with a gain $\geq 0.5$ (the gray dashed line represents this limit). The plot shows in blue the NC gain profile, corresponding to the primary beam attenuation, over the entire observable interval, while the dashed yellow line represent the net observed window that, for this target, is stopped 480 s in advance due to schedule constraints.}
    \label{fig:Gain}
\end{figure}

Using the radiometer equation \citep{Lorimer_and_Kramer}, as in \citet{NC_Project_4}, we can estimate the flux density of a single burst detected with the NC with a given $S/N$ as:
\begin{equation} \label{Eq: Flux}
    F_\nu = S/N \; \frac{\rm{SEFD}}{A \sqrt{N_{\rm{p}} N_{\rm{c}} (1-\xi) \Delta\nu_{\rm{ch}} w}} \frac{1}{\mathcal{G}(\rm{ToA})}.
\end{equation}
The System Equivalent Flux Density (SEFD) is SEFD $= 8400$ Jy \citep{NC_Project_2} for every receiver and the number of receivers is $A = 64$, in the current 16 cylinders configuration. The number of polarizations in our case is $N_{\rm{p}}=1$, due to the fact that the NC has the dipoles aligned along the focal direction. The burst full width at half maximum (FWHM) is $w$, the number of frequency channels is $N_{\rm{c}} = 1024$ each of width $\Delta\nu_{\rm{ch}} = 14.468$ kHz, while $\xi$ (usually $\lesssim 5\%$) is the fraction of channels, with respect to the total, that are excluded due to the presence of radio frequency interferences (RFIs). The gain of the NC is represented by $\mathcal{G}(\rm{ToA}) \in [0.493,1] $ and is linked to the primary beam attenuation of each single antenna element. The latter depends on the time of arrival ($\rm{ToA}$) of the burst, the attenuation is minimum for bursts arriving at the time of the peak of the gain and it increases away from it (see Fig. \ref{fig:Gain}).
In fact, for the NC transit telescope, the observed flux density of a celestial source changes as it moves across the telescope field of view, according to the profile shown in Fig. \ref{fig:Gain}. This beam-forming response is fitted from the telescope gain profile and applied to the magnetar analysis of Sec. \ref{sec:methods}. Differently from previous works \citep{NC_Project_1,NC_Project_2,NC_Project_3}, but coherently with \cite{NC_Project_4}, here we are using a dense source tracking, with updates of the beam-steering coefficients that occur every 5 s of observation.
Furthermore, the North hemisphere location and the mechanical steering properties of the NC allow us to observe only sources with sky declination $\gtrsim 0 \degree$.
As our targets lie in the $-0.996\degree < b < 5.119\degree$ range, Galactic plane emission is expected to increase the telescope SEFD. We used the 408 MHz all-sky map provided by \cite{Map_408MHz_2014} to estimate the sky temperature in a 6$\degree$ area centered on each magnetar of the sample. We found that the sky contribution to the SEFD is negligible (less than 3\%) for any magnetar of our sample.

\subsection{Data analysis strategy}
The observations have been stored as 16-bit SIGPROC filterbank format data \citep{SIGPROC_Lorimer}.
The narrow bandwidth of the NC and the low presence of RFIs allowed us to analyze most of the filterbanks without using RFI cleaning tools. Anyway, when the individual filterbanks provided a high number of candidates due to RFI, we have used the Inter-Quartile Range Mitigation (\texttt{IQRM})\footnote{See \href{https://gitlab.com/kmrajwade/iqrm\_apollo}{\texttt{gitlab.com/kmrajwade/iqrm\_apollo}}.} real-time adaptive RFI masking \citep{IQRM}.
The incoherent dedispersion and the search for radio transients have been performed using the \texttt{HEIMDALL}\footnote{See \href{https://sourceforge.net/p/heimdall-astro/wiki/Home/}{\texttt{sourceforge.net/p/heimdall-astro/wiki/Home/}}.} algorithm \citep{Heimdall}.
Taking into account the high uncertainties on the Galactic DM contribution (see Table \ref{tab:Magnetars_info}), due to not well known distances and different possible models, we searched for candidates over a wide DM range, from 20 to 3000 pc cm$^{-3}$. Only in the case of SGR\,J1935+2154, we used a narrower DM range, between 300 to 360 pc cm$^{-3}$, centered around the known $\rm{DM}_{\rm{J1935}} \sim 332.7$ pc cm$^{-3}$ \citep{SGR1935_CHIME}.
Our best use of our computing capabilities allowed us to perform a search adopting a fine DM sampling, with $\sim 25000$ DM-steps ($\sim 1100$ in the case of SGR\,J1935+2154) and looking for bursts with a maximum width of $\sim70$ ms.
The \texttt{HEIMDALL} generated candidates with signal-to-noise ratio greater than 7 have been classified with \texttt{FETCH}\footnote{See \href{https://github.com/devanshkv/fetch}{\texttt{github.com/devanshkv/fetch}}.}, a convolutional neural network trained to distinguish between RFIs and astrophysical dedispersed signals \citep{FETCH}.
We tested the efficiency of \texttt{FETCH} models on NC data and we found that combining the probabilities of the different models improve the detectability of dispersed signals.
Finally, every candidate, that passed all the selection steps, undergoes a deep investigation with a human inspection.

\subsection{Observing campaign}
\begin{figure*}[h]
    \centering
    \includegraphics[width=\linewidth]{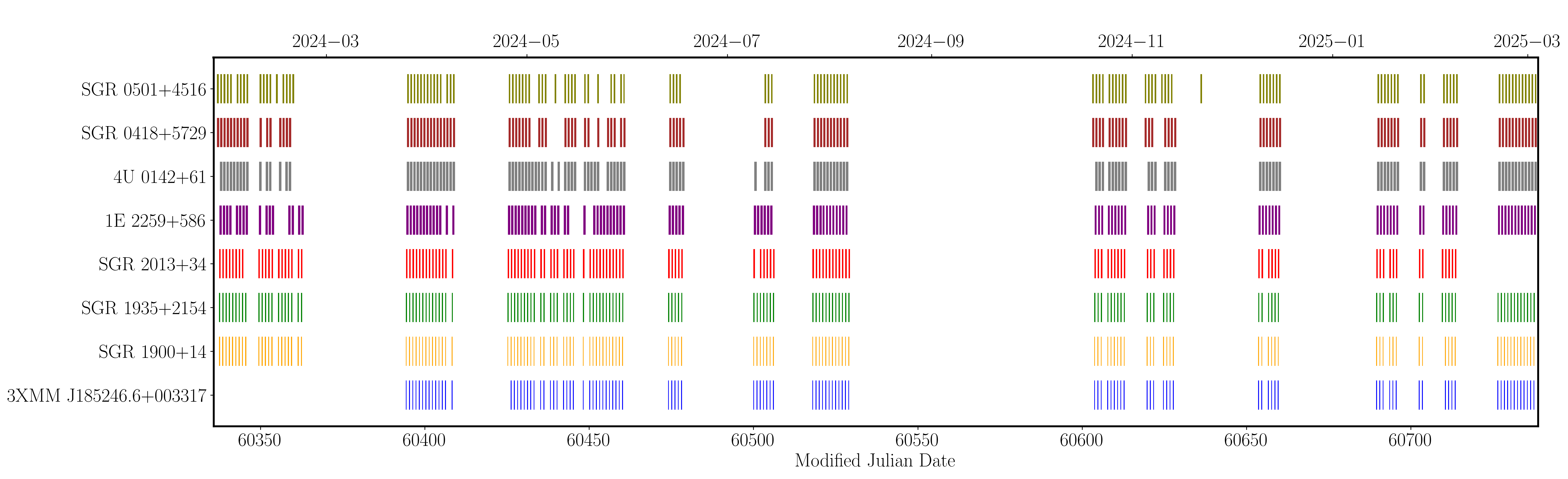}
    \caption{Observing windows during the monitoring campaign. The total exposure is 565.92\;h. Each vertical notch represents an observation, the typical duration of daily exposure for the different sources can be seen in Fig. \ref{fig:Dayobs}.}
    \label{fig:Observing_win}
\end{figure*}

\begin{figure}[h]
    \centering
    \includegraphics[width=\linewidth]{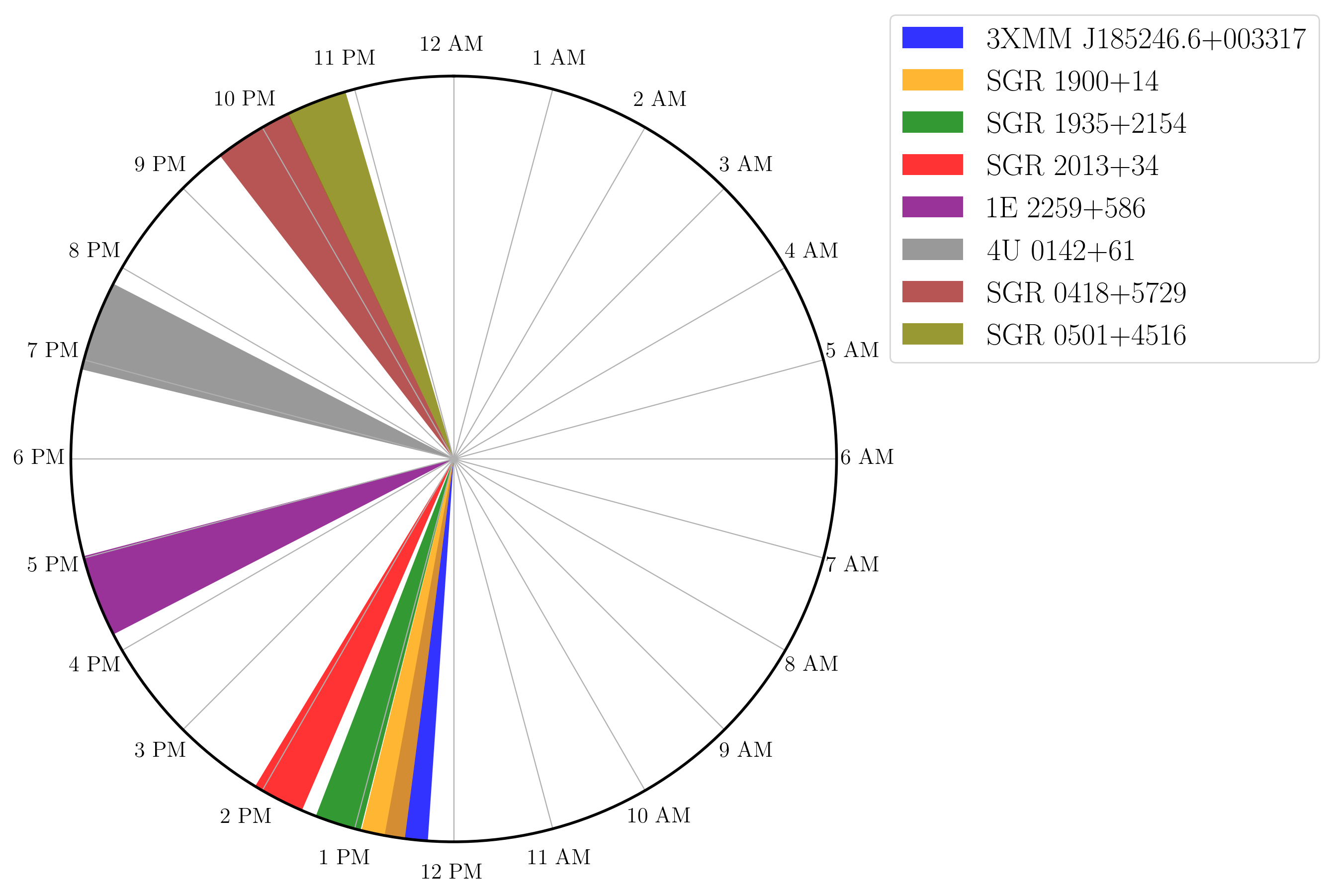}
    \caption{The figure shows an example of a daily time distribution of the observable windows of the targets. The NC observing intervals on each target shift by $\sim 4$ minutes per day in advance.}
    \label{fig:Dayobs}
\end{figure}

We observed the targets from 2024-01-27 to 2025-03-03 (see Fig. \ref{fig:Observing_win}) with daily observations as allowed by the telescope scheduling constraints.
In Table \ref{tab:Magnetars_results}, we report the amount of observing time  for each target.
Most of the observed magnetars are distributed in the Galactic plane, therefore the RA distribution of the targets is such that there is often overlap of visibility from the NC (see Fig.\ref{fig:Dayobs}).
Taking into account the overlaps of the targets and the antennas movement delays, we have been forced to cut some of the observing intervals at the extremes. The crop of the observing windows reduces the amount of on-source time for the NC. However, this cut of the extremes of the interval implies that the parts that are erased are the ones that are less sensitive due to primary beam attenuation (as shown in Fig.\ref{fig:Gain}) and we took this into account in the results of our work.

\section{Methods}\label{sec:methods}
If we apply Eq. \ref{Eq: Flux} to our observing campaign, considering only candidates with $S/N$ $ \geq 7$, the NC minimum detectable flux density is $F_\nu \sim 7.55 \;  \left( \frac{w}{\rm{[ms]}} \right)^{-1/2} \frac{1}{\mathcal{G}(\rm{ToA})}$ Jy.
If we convert the flux density into an energy for a source located at distance $D$, using the NC bandwidth $\Delta \nu = 14.815$ MHz and assuming a pulse duration of $w = 1$ ms, we find the highest ($\mathcal{G}(\rm{ToA}) = 0.493$) and lowest ($\mathcal{G}(\rm{ToA}) = 1$) detectable energies:
\begin{equation}\label{EQ:E_min}
    E_{\rm{min}} = (1.34 - 2.72) \times 10^{20} \left( \frac{D}{\rm{[pc]}} \right)^{2} \rm{erg}.
\end{equation}
The intervals of the minimum detectable energies for each magnetar are listed in the third column of Table \ref{tab:Magnetars_results}. The uncertainties on the distances of the targets (listed in Table \ref{tab:Magnetars_info}) can influence the minimum energies estimates, but even assuming the extreme values, we verified that the final conclusions of this work are not significantly influenced by these changes.

Our sensitivity threshold corresponds to a spectral luminosity of $\sim 7.3 \times 10^{22}$ erg s$^{-1}$ Hz$^{-1}$ for a 1 ms burst assuming $D = 2$ kpc (i.e. the minimum distance in our sample of magnetars). On the other hand, using the maximum distance among our targets, $D = 12.5$ kpc, the limit grows to $\sim 2.8 \times 10^{24}$ erg s$^{-1}$ Hz$^{-1}$.
Comparing our thresholds with the known population of fast ($\lesssim$1 s) radio transients \citep{Nimmo_2022_frb}, we expect to be sensitive to any FRB-like event from our sample of magnetars. Furthermore, especially for the closest magnetars, the NC has the sensitivity to detect events with giant-pulse energy \citep{Giant_pulse_karu} and the most energetic single pulses by rotating radio transients and `ordinary' pulsars \citep{RRATS_pop}.

\begin{table*}
    \caption{Inferred upper limits on the magnetars sample.}
    \resizebox{\textwidth}{!}{
    \renewcommand{\arraystretch}{1.5}
    \begin{tabular}{l c c c c c c c}
        \hline
        \hline
        Name & Total exposure & E$_{\rm{min}}$ & $\mathcal{R} \left( \rm{E} > \rm{E}_{\rm{min}} \right)$ & $\mathcal{R} \left( \rm{E} > \rm{E}_{\rm{J1935}} \right) $, $\gamma=1$ & $\mathcal{R} \left( \rm{E} > \rm{E}_{\rm{J1935}} \right) $, $\gamma=1.3$ & $\mathcal{R} \left( \rm{E} > \rm{E}_{\rm{J1935}} \right) $, $\gamma=1.6$\\
        & (hours) & $10^{26}$ (erg) & (yr$^{-1}$) & (yr$^{-1}$) & (yr$^{-1}$) & (yr$^{-1}$) \\
        \hline
        3XMM\,J185246.6+003317 & 33.46 & $67-137$ & < 785 & < 400 & < 8.78 & < 0.10 \\

        SGR\,1900+14 & 40.58 & $209-424$ & < 647 & < 342 & < 10.19 & < 0.17 \\

        SGR\,J1935+2154 & 52.34 & $108-220$ & < 502 & < 260 & < 6.48 & < 0.09 \\

        SGR\,2013+34 & 63.38 & $104-210$ & < 414 & < 214 & < 5.36 & < 0.07 \\

        1E\,2259+586 & 97.03 & $14-28$ & < 271 & < 131 & < 1.91 & < 0.01 \\

        4U\,0142+61 & 116.08 & $17-35$ & < 226 & < 110 & < 1.71 & < 0.01 \\

        SGR\,0418+5729 & 92.95 & $5-11$ & < 283 & < 133 & < 1.49 & < 0.01 \\

        SGR\,0501+4516 & 70.10 & $5-11$ & < 375 & < 176 & < 1.97 & < 0.01 \\
        
        \hline
        
        All magnetars & 565.92 & $5-424$ & < 46 & < 2.92 & < 0.05 & < $4.04 \times 10^{-4}$ \\

        All magnetars (NO SGR\,2013+34) & 502.54 & $5-424$ & < 52 & < 3.74 & < 0.06 & < $4.76 \times 10^{-4}$ \\
        
    \end{tabular}}\\ \\
    \textbf{Notes.} The second column contains the amount of hours observed on each target. The third column represents the interval of the minimum detectable energy above which we defined the upper limit on the rate of events. The fourth column shows the Poissonian upper limit at 95\% confidence level on the observable burst rate. The last three columns list the inferred upper limits on the rates of events with an energy grater than the FRB-like event of the magnetar SGR\,J1935+2154 using Eq. \ref{EQ:E_rate_3} for 3 different power-law indexes (as indicated in each column).  \\
    \label{tab:Magnetars_results}
\end{table*}

\subsection{Power law energy emission model} \label{subsec:PL-model}
Similarly to \cite{NC_Project_3}, assuming an energy power-law distribution of single bursts, we defined the rate of events above the minimum detectable energy coming from a single magnetar in the primary beam attenuation interval, $j$, as:
\begin{equation}\label{EQ:E_rate_1}
    \mathcal{R} \left(\lambda_{\rm{mag}},\gamma ; \; E > E_{\rm{min,j}} \right) = \int_{E_{\rm{min,j}}}^{E_{\rm{max}}} K_0 \; \left( \frac{E}{E_{\rm{0}}} \right)^{-\gamma} \; dE.
\end{equation}
Differently from \cite{NC_Project_3}, we have not restricted the rate of events to the ones with energy higher than $E_0$ (i.e. using the Heaviside function $\Theta[E-E_0]$), in fact, in our work the low energy events strongly influence the expected bursts rate.
Looking at Eq. \ref{EQ:E_rate_1}, we fitted the NC gain profile (Fig. \ref{fig:Gain}) by dividing it in short intervals, therefore, $E_{\rm{min,j}}$ is the NC sensitivity, in the gain interval j, computed with Eq. \ref{Eq: Flux}.
The maximum energy of the distribution is $E_{\rm{max}}$, while $E_0$ is a reference energy used for the normalization and $K$ is a normalization constant that ensures:
\begin{equation}\label{EQ:Norm}
    \lambda_{\rm{mag}} = \int_{E_0}^{E_{\rm{max}}} K_0 \; \left( \frac{E}{E_0} \right)^{-\gamma} \; dE.
\end{equation}
The latter equation defines the number of events emitted by a magnetar with energy above $E_0$.
Crucially, the free parameters of our model are $\lambda_{\rm{mag}}$ and the power-law index $\gamma$.
Similarly to \cite{NC_Project_3}, we adopted a reference energy $E_0 = 3.2 \times 10^{34}$ erg, which represents the energy of the FBR-like event of SGR\,J1935+2154 \citep{Margalit_2020} considering our 9 kpc assumption of the distance of SGR\,J1935+2154 (\cite{NC_Project_3} assumed $E_0 = 2 \times 10^{34}$ erg due to different distance estimate).
Furthermore, we assume that the maximum energy, that a magnetar is able to emit, depends on the energy reservoir of the neutron star that is function of the surface magnetic field strength \citep{Margalit_2020}:
\begin{equation}\label{EQ:Emax}
    E_{\rm{max}} = E_{\rm{mag}} \times \eta \simeq 3 \times 10^{44} \left( \frac{B}{10^{16} \; \rm{G}} \right)^2 \left( \frac{\eta}{10^{-5}} \right)\; \rm{erg} .
\end{equation}
Taking into account the uncertainties on the radio efficiency and the complex magnetic field structure of magnetars, we used a common order of magnitude for the maximum energy for all the magnetars: $E_{\rm{max}} = 1.2 \times 10^{41}$ erg. Here we assumed a surface magnetic field of $B_{\rm{dip}} = 2 \times 10^{14}$ G and an efficiency of the radio emission $\eta = 10^{-5}$ for each magnetar, as in the case of SGR\,J1935+2154 radio bursts \citep{Margalit_2020}.

Combining the normalization of Eq. \ref{EQ:Norm} with the rate of events of Eq. \ref{EQ:E_rate_1}, we obtain:
\begin{equation}\label{EQ:E_rate_2}
    \mathcal{R} \left(\lambda_{\rm{mag}},\gamma ; \; E > E_{\rm{min,j}} \right) = \lambda_{\rm{mag}} \frac{E_{\rm{max}}^{\;\;\;\;1-\gamma} - E_{\rm{min,j}}^{\;\;\;\;1-\gamma}}{E_{\rm{max}}^{\;\;\;\;1-\gamma} - E_{\rm{0}}^{\;1-\gamma}}.
\end{equation}
From the latter equation, knowing that in our case $E_{\rm{max}} \gg E_0 \gg E_{\rm{min}}$, we can see that the choice of the maximum energy influences the inferred rate of events only when $\gamma \simeq 1$.
If we perform a sum weighted on the time spent in every primary beam attenuation interval $T_{\rm{j}}$, we obtain:
\begin{equation}\label{EQ:E_rate_3}
    \mathcal{R} \left(\lambda_{\rm{mag}},\gamma \right) = \lambda_{\rm{mag}} \frac{\sum_{\rm{j}}\left[\left(E_{\rm{max}}^{\;\;\;\;1-\gamma} - E_{\rm{min,j}}^{\;\;\;\;1-\gamma}\right) \cdot T_{\rm{j}}\right]}{\left( E_{\rm{max}}^{\;\;\;\;1-\gamma} - E_{\rm{0}}^{\;1-\gamma} \right) \cdot \sum_{\rm{i}} \left[ T_{\rm{j}}\right]}.
\end{equation}
Finally, in the case of the entire sample of magnetars, the rate of events expected from our sample is the sum of the rates of the single magnetars:
\begin{equation}\label{EQ:E_rate_4}
    \mathcal{R}_{\rm{tot}} \left(\lambda_{\rm{mag}},\gamma \right) = \sum_{i=1}^{N_{\rm{mag}}} \mathcal{R}_{\rm{i}}.
\end{equation}

\section{Results and discussion} \label{sec:results}
No magnetar-related detections were found during the monitoring campaign.
We exploited the observing hours to set upper limits on the burst rate of impulsive events from each target in our sample of Galactic magnetars. We assumed a Poisson distribution of events in time and the inferred upper limits for each magnetar are listed in the fourth column of Table \ref{tab:Magnetars_results}.
Considering the entire $\sim$560 hours monitoring, the upper limit on the burst rate at a 95$\%$ confidence level is $\mathcal{R}_{\rm{tot}} < 46$ yr$^{-1}$ \citep[][Table 1]{Gehrels_1986}. If we exclude the magnetar candidate SGR\,2013+34, the `clean' upper limit grows up to $\mathcal{R}_{\rm{tot,clean}} < 52$ yr$^{-1}$.
The upper limits refer only to bursts with an energy greater than the minimum detectable energy for each magnetar.
In fact, each target has a different minimum detectable energy interval and a different upper limit on the burst rate, related to the amount of on-source time.
The result of our power-law energy distribution model (see Sec. \ref{sec:methods}) is a rate of single pulse events and it can be represented using a colored scale that shows the variations of the number of detectable events in the parameter space.
The rate $\mathcal{R}_{\rm{tot}}$, is function of the free parameters of the model, that are the slope of the power-law energy distribution ($\gamma$) and the number of events with energy greater than the energy emitted by the FRB-like event of SGR\,J1935+2154 ($\lambda_{\rm{mag}}$).
If we combine the number of observable events, computed with the method described before (look at \ref{subsec:PL-model}), with the upper limits obtained by our observations, we can investigate the parameter space of our model and infer which are the favored combinations of the free parameters.
The resulting plots, for the entire magnetars sample, are showed in Fig. \ref{fig:Rate_mag}, while the plots for each magnetar are reported in appendix \ref{Appendix:single_plots} for completeness.

\begin{figure*}
    \centering
        \includegraphics[width=0.49\linewidth]{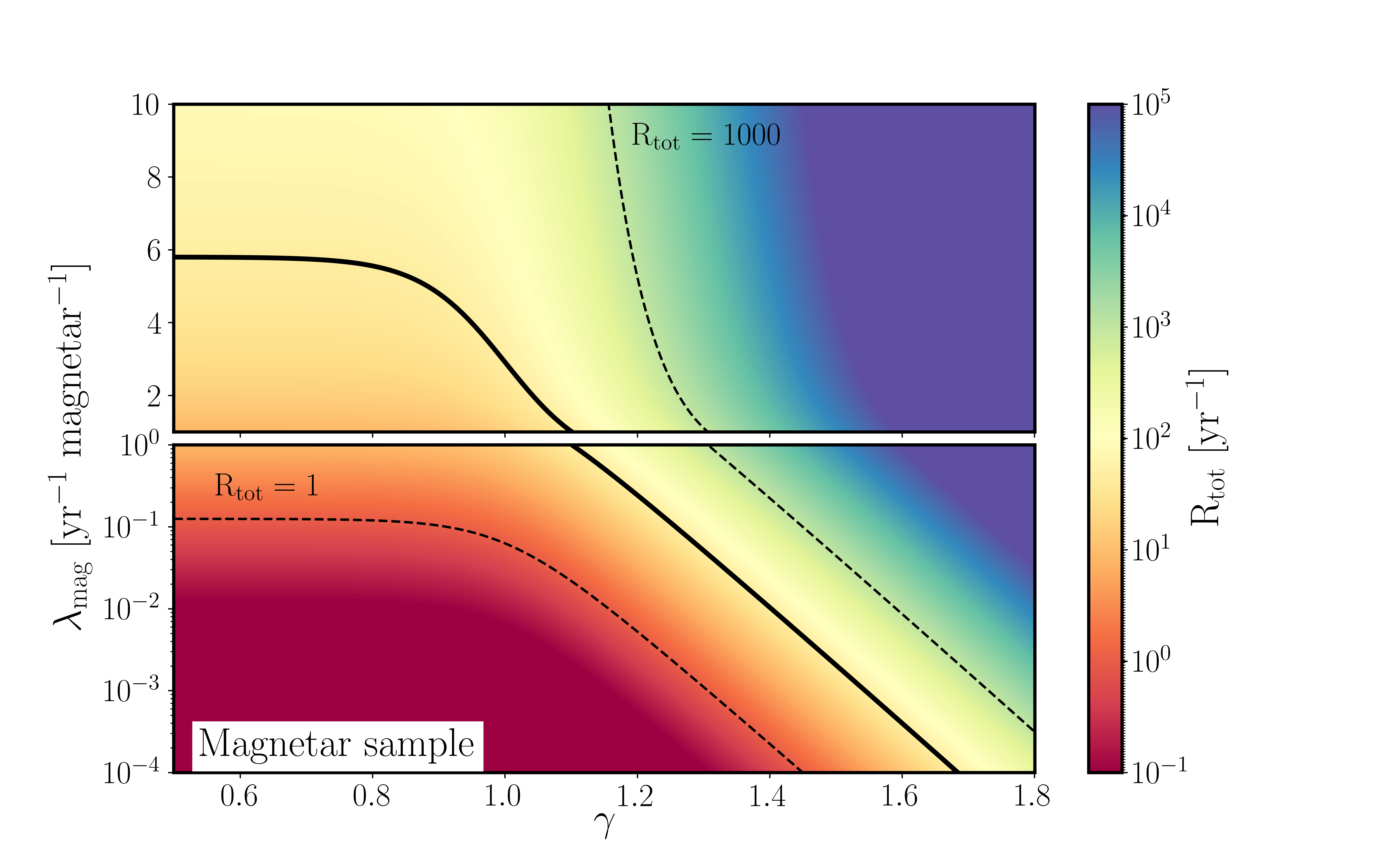}
        \includegraphics[width=0.49\linewidth]{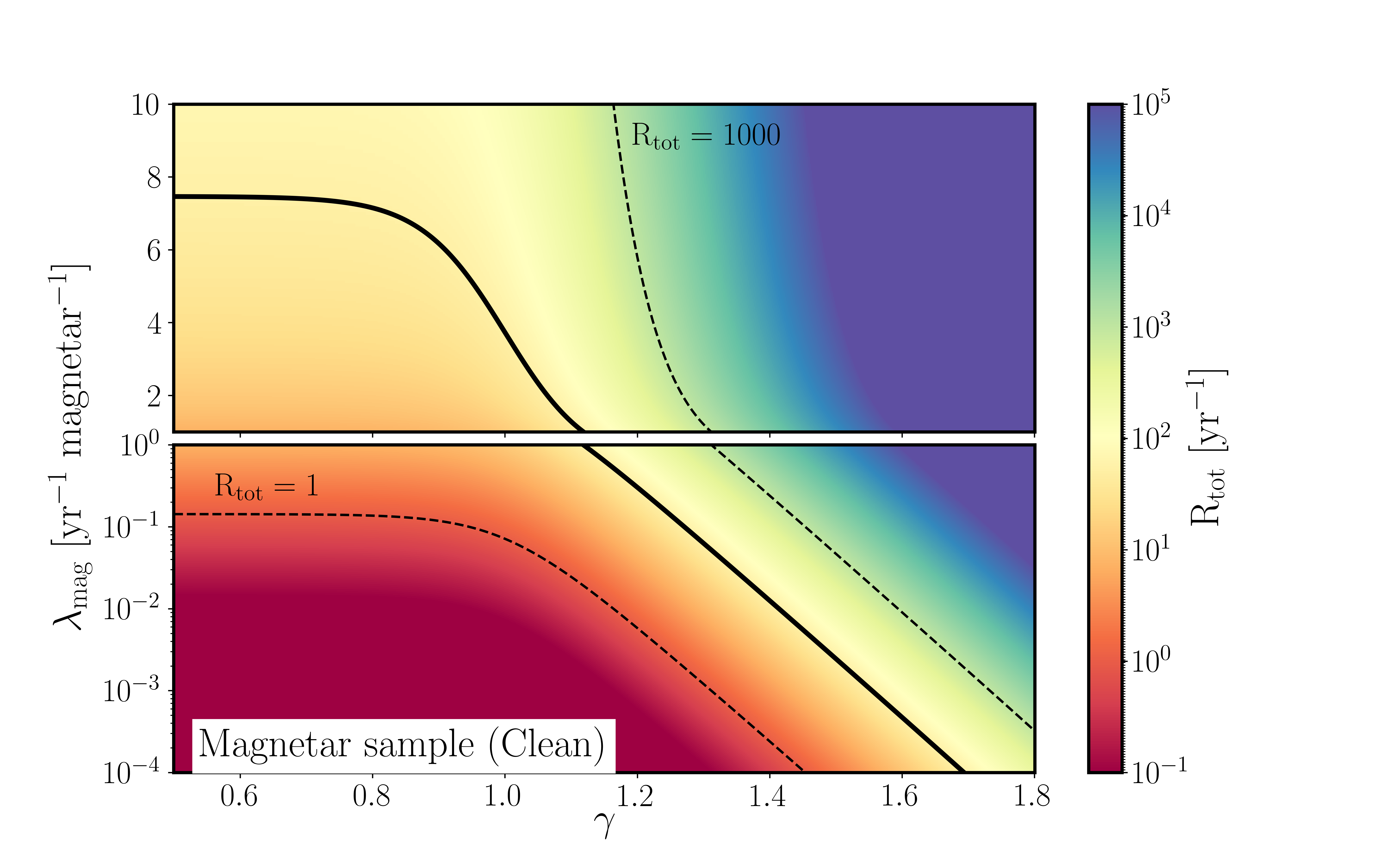}

    \caption{The plots show a graphical representation of our power-law energy distribution model, for both the entire sample of targets (left) and excluding the magnetar candidate SGR\,2013+34 (right). The x-axis represents the slope of the power-law energy distribution $\gamma$, while on the y-axis there is the number of events with energy greater than the energy emitted by the FRB-like event of SGR\,J1935+2154 $\lambda_{\rm{mag}}$. The latter are the two free parameters of the model and their combinations produce $\mathcal{R}_{\rm{tot}}$ (Eq. \ref{EQ:E_rate_4}) for the entire magnetars sample, respectively.
    The rate of events obtained in every case, are showed using a fixed logarithmic color code presented in the colorbars.
    The solid black lines represent the upper limits, obtained with our observing campaign, on the possible combinations of $\gamma$ and $\lambda_{\rm{mag}}$, while the black dashed lines show reference values for graphical purposes. The difference between upper and lower panels is the scale of the y-axis, the upper panels are in linear scale and the lower panels in logarithmic scale.}
    \label{fig:Rate_mag}
\end{figure*} 

In our model, the rate of high-energy events $\lambda_{\rm{mag}}$ has the role of a normalization, therefore, having a flat energy distribution (i.e. low $\gamma$) implies that we do not expect a large population of low-energy detectable events.
On the other hand, rapidly decreasing power-law distributions (i.e. high $\gamma$) produce a great number of low-energy events that we expect to detect.
Looking at Fig. \ref{fig:Rate_mag}, our results combined with the observations upper limits, suggest that flatter energy distributions are favored with respect to steeper ones.
We can notice that, values of $\gamma < 1$ produce a flat upper limit in our plots, since in that case, the $E_{\rm{max}}$ contribution dominates in Eq. \ref{EQ:E_rate_2}.

\subsection{Comparison with FRB works}
Since the SGR\,J1935+2154 FRB-like event in 2020 \citep{SGR1935_CHIME}, a considerable effort has been made to investigate the possible link between magnetars and FRBs.
Works as \cite{SGR1935_CHIME,NC_Project_3,SGR1935_Kirsten} already set upper and lower limits on the expected burst rate of FRB-like signals from magnetars, while \cite{FAST_1935_3XMM} set deep upper limits on the energy emitted by the magnetars SGR\,J1935+2154 and 3XMM\,J185246.6+003317 during quiescent phases.

\cite{SGR1935_Kirsten} used few European radio telescopes to monitor the activity of SGR\,J1935+2154, obtaining two detections at 1.4 GHz in a total of 763.3 hours. Knowing that the time separation between the two detected bursts was only $\sim 1.4$ s, they fitted the time intervals between subsequent bursts with a Weibull distribution.
On the other hand, in our case of non detections, we assumed a Poissonian distribution of events as explained earlier.
If we apply our statistics to the \cite{SGR1935_Kirsten} work, ignoring the statistical differences and knowing that both the observing frequency bands and sensitivities are different, we obtain an expected number of detectable events from magnetars between 4 - 34 yr$^{-1}$, that is consistent with our upper limits.

\begin{figure*}
    \centering
        \includegraphics[width=0.49\linewidth]{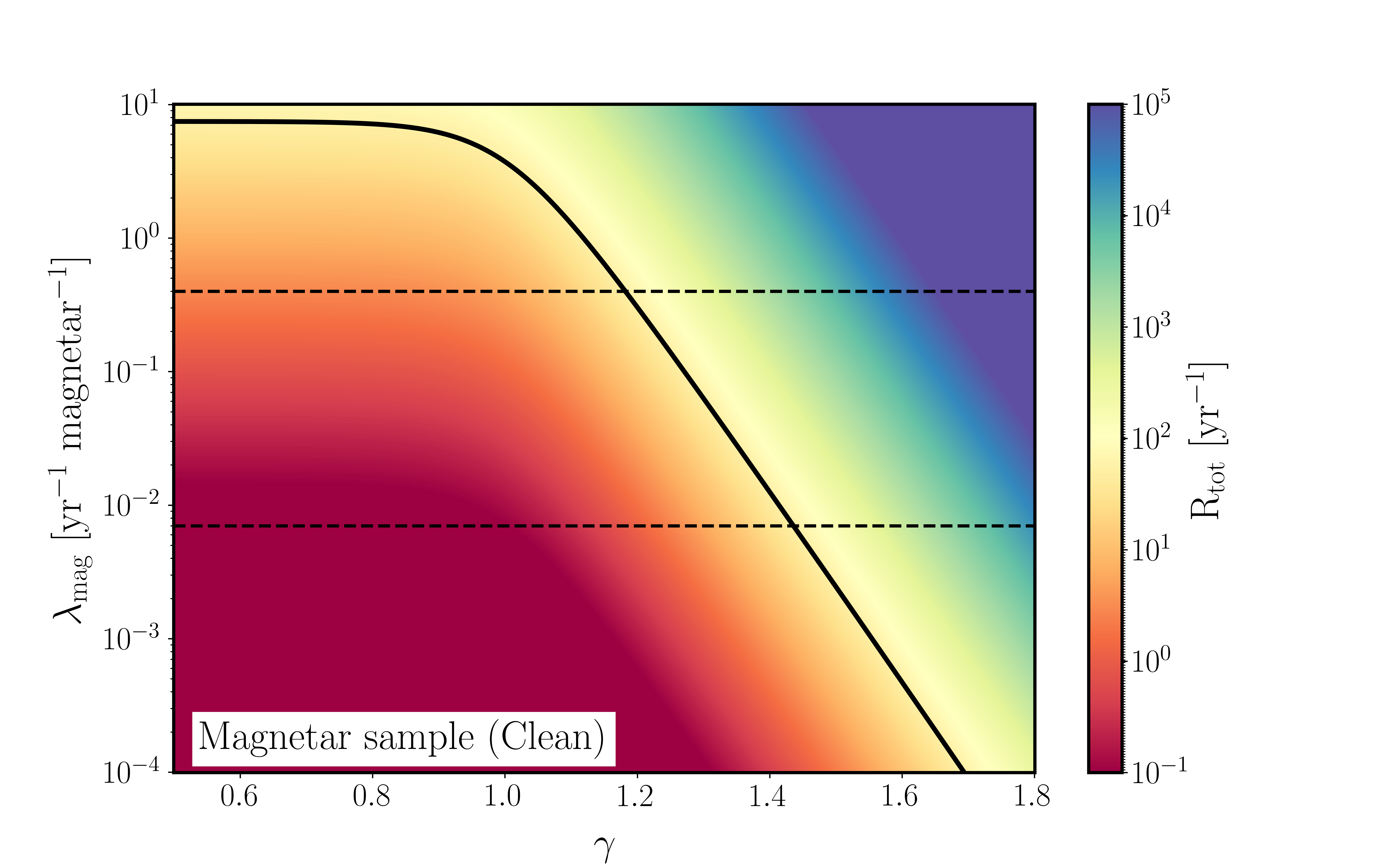}
        \includegraphics[width=0.49\linewidth]{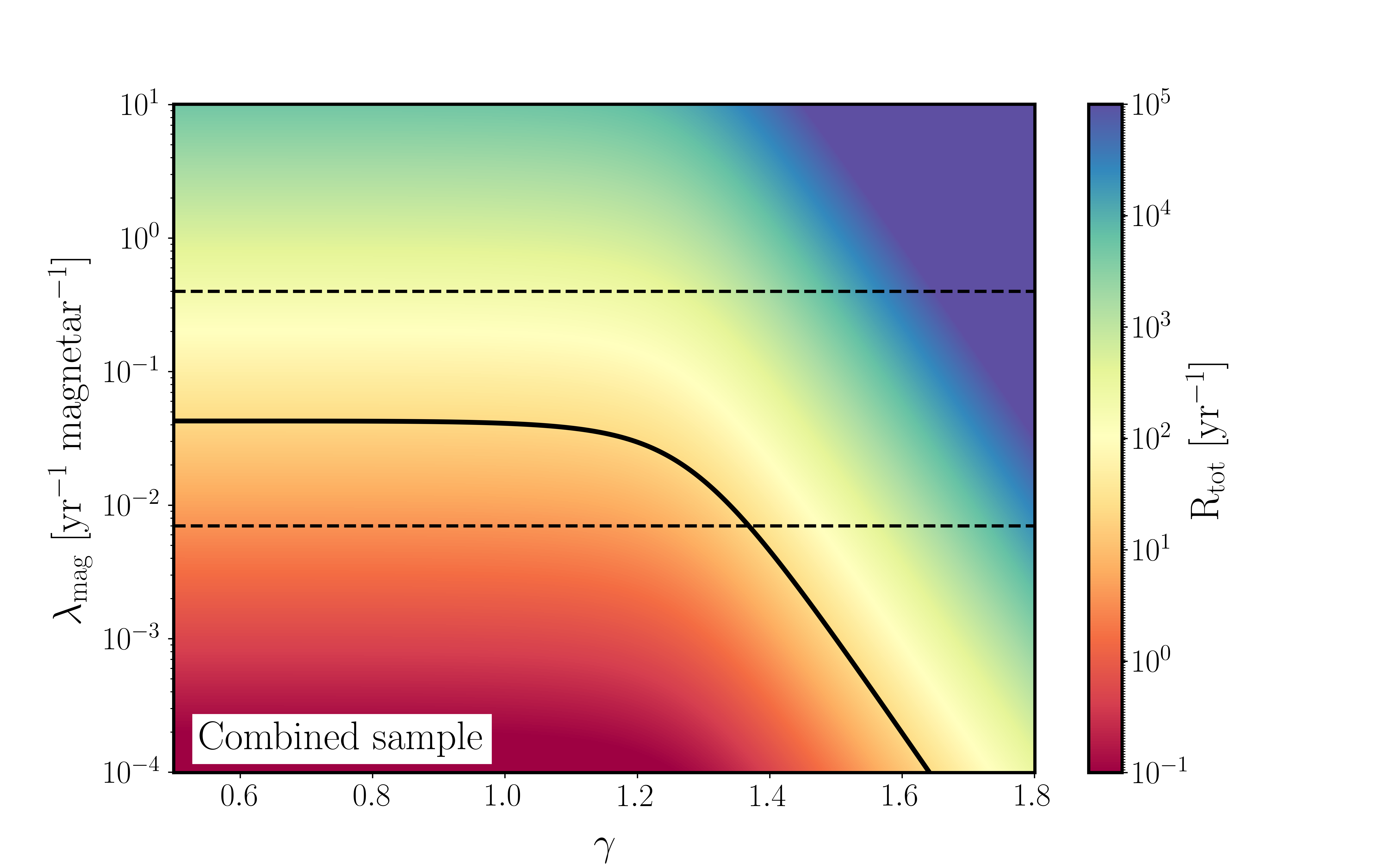}
    \caption{The figure shows a graphical representation of our power-law distribution model, same as the right panel of figure \ref{fig:Rate_mag}.
    The dashed lines represent the \cite{SGR1935_CHIME} lower and upper limits at 95 \% confidence level on the rate of events SGR\,J1935+2154-like.
    We also adopt a different scale for a better visualization.
    In the right panel, we combined our campaign with the nearby galaxies observations of \cite{NC_Project_3}.
    }
    \label{fig:Rate_mag_comb}
\end{figure*}

If we take into account literature observations on extragalactic targets, \cite{SGR1935_CHIME} observed SGR\,J1935+2154 and 15 nearby galaxies setting constrains on the burst rate of FRB-like events. They assumed a number of magnetars in the Milky Way (MW) $N_{\rm{mag}} \sim 30$ and that the number of magnetars in a galaxy is directly related to its star formation rate, resulting in a rate of SGR\,J1935+2154-like bursts per magnetar of $0.007 < \lambda_{\rm{mag}} < 0.4$ yr$^{-1}$.
A similar reasoning was followed in \cite{NC_Project_3}, from which a consistent upper limit of $\lambda_{\rm{mag}} < 0.42$ yr$^{-1}$ was obtained.
We can use the constraints on $\lambda_{\rm{mag}}$ from \cite{SGR1935_CHIME}, to investigate which combinations of our parameter space are favored by our upper limits. The resulting plot is shown in the left panel of Fig. \ref{fig:Rate_mag_comb}, where we can see that only power-law indexes $\gamma\lesssim1.4$ produce observable rates consistent with the constraints set by \cite{SGR1935_CHIME}.
Furthermore, we can exploit the $\sim 695$ hours observed in the nearby galaxies campaign of \cite{NC_Project_3} obtaining an improved NC upper limit on the burst rate resulting in $\lambda_{\rm{mag}}<0.043$ yr$^{-1}$.
The resulting plot, compared with the literature upper limits, is shown in the right panel Fig. \ref{fig:Rate_mag_comb}.
We used the same conservative assumptions of \cite{SGR1935_CHIME,NC_Project_3} by setting the number of magnetars in the MW as $N_{\rm{mag}} = 29$, but exploiting the low-energy events (i.e. removing the Heaviside function in Eq. 4 of \cite{NC_Project_3}).
Crucially, $N_{\rm{mag}} = 29$ is the number of known magnetars in our Galaxy, knowing that many more magnetars, that we have not yet discovered, may actually be present in the MW.
If we define magnetars as sources that can emit an SGR-like flares, works as (\citealt{Mag_pop_Muno,Mag_pop_Gullon}; Ronchi et al., in preparation) suggest that the number of currently active neutron stars in the MW, that could behave as magnetars, is between 500 and 800.
Taking into account this estimate, the upper limit (solid black line) showed in the right panel of Fig. \ref{fig:Rate_mag_comb} is most likely overestimated.
In our case, we can notice that, if we adopt a value of $N_{\rm{mag}}=500$ consistent with these population studies, we find a Northern Cross upper limit $\lambda_{\rm{mag}} < 0.0025$ yr$^{-1}$.

\section{Summary and conclusions}\label{sec:conclusions}

In this work, we used the Northern Cross radio telescope to monitor seven known Galactic magnetars that are visible from the site (Tab \ref{tab:Magnetars_info}).
Our $\sim$560 hours of observations did not produce any sound detection of events directly linked to magnetars.
We exploited our long-term monitoring campaign to constrain the rate of single pulses from this class of objects, making also considerations on the occurrence of FRB-like events.
Our upper limit of $<$52 yr$^{-1}$ at 95\% confidence level on the cleaned sample of magnetars is consistent with the ones estimated in previous observations \citep{SGR1935_Kirsten}.
Furthermore, we modeled the single pulses emission from magnetars using a power law energy distribution, similarly to \cite{NC_Project_3}.
Our upper limits constrain the free parameters of the model ($\gamma$ and $\lambda_{\rm{mag}}$), suggesting a flat energy distribution of events (see Table \ref{tab:Magnetars_results} and Fig. \ref{fig:Rate_mag}).
If we include nearby galaxies observations \citep{SGR1935_CHIME,NC_Project_3}, the constraints on the rate become more stringent and the flatter energy distributions are even more preferred (see Fig. \ref{fig:Rate_mag_comb}).
Furthermore, adopting a more realistic estimate of the number of active magnetars in MW-like galaxies ($N_{\rm{mag}}=500$) the limits became even more stringent.

The fact that, during our monitoring campaign, none of the targets showed X-ray bursting activity or outburst phases could be a hint that FRB-like events are indeed related to magnetar-like activity windows, as in the case of SGR\,J1935+2154 and of their radio emission in general.
As already discussed, the typical inferred luminosity of extragalactic FRBs is few order of magnitude higher than the FRB-like event observed from the magnetar SGR\,J1935+2154.

It is relevant to point out that our observations constrain $\lambda_{\rm mag}$ to be $\lambda_{\rm mag} < 7.4$ yr$^{-1}$ for flat values of the energy slope $\gamma$, that is, $\gamma \lesssim 1.1$, while for steeper energy slopes, i.e. for $\gamma \gtrsim 1.4$, the chance of observing an FRB-like event with energy greater than that of SGR\,J1935+2154 becomes increasingly small. Furthermore, if we combine our observations with nearby galaxies observations \citep{SGR1935_CHIME,NC_Project_3} the constrain on $\lambda_{\rm mag}$ for the flat energy slopes becomes $0.007 < \lambda_{\rm mag} < 0.043$ yr$^{-1}$.

These results are somewhat in tension with models that attempt to explain the all-sky FRB rate with radio bursts from magnetars similar to SGR\,J1935+2154.

Indeed, \citet{James_FRBpop} found that the extragalactic FRB rate can be explained in terms of evolution of the cosmic star formation rate, with a slope of the luminosity distribution $\gamma \simeq 2.1$.\footnote{The value is converted from their cumulative distribution to our probability distribution.} If there is no significant evolution of the slope with redshift, this result is disfavored by our observations that tend to exclude slopes steeper than $\sim$1.5.

Furthermore, \cite{Margalit_2020}, based on radio observations of SGR\,J1935+2154 \citep{SGR1935_CHIME}, assumed $0.0036 < \lambda_{\rm mag} < 0.8$ yr$^{-1}$ and found that the slope of the luminosity distribution needs to be in the $1 < \gamma < 1.5$ range to fit the extragalactic FRB rate. For this reason, they speculate on the existence of an additional population of more exotic magnetars, not generated by core-collapsed supernovae, needed to be included in the model, as the event rate of SGR\,J1935+2154-like magnetars is too small to account for the observed extragalactic rate. As our results reduce the ranges of allowed $\lambda_{\rm mag}$ and $\gamma$ values, it implies a larger contribution from this possible population of exotic magnetars.

\begin{acknowledgements}
The reported data were collected during the phase of the INAF scientific exploitation with the NC radio telescope. We thank the referee, for their comments and suggestions that greatly improved the paper, and Nanda Rea, for useful discussion. This publication  was produced while AG was attending the PhD program in Space Science and Technology at the University of Trento, Cycle XXXIX, with the support of a scholarship financed by the Ministerial Decree no. 118 of 2nd March 2023, based on the NRRP - funded by the European Union - NextGenerationEU - Mission 4 "Education and Research", Component 1 "Enhancement of the offer of educational services: from nurseries to universities” - Investment 4.1 “Extension of the number of research doctorates and innovative doctorates for public administration and cultural heritage” -- CUP E66E23000110001. Part of the research activities described in this paper were carried out with contribution of the NextGenerationEU funds within the National Recovery and Resilience Plan (PNRR), Mission 4 - Education and Research, Component 2 - From Research to Business (M4C2), Investment Line 3.1 - Strengthening and creation of Research Infrastructures, Project IR0000026 – Next Generation Croce del Nord.
\end{acknowledgements}

\bibliography{Biblio}{}

\begin{thebibliography}{105}
\expandafter\ifx\csname natexlab\endcsname\relax\def\natexlab#1{#1}\fi

\bibitem[{{Agarwal} {et~al.}(2020){Agarwal}, {Aggarwal}, {Burke-Spolaor},
  {Lorimer}, \& {Garver-Daniels}}]{FETCH}
{Agarwal}, D., {Aggarwal}, K., {Burke-Spolaor}, S., {Lorimer}, D.~R., \&
  {Garver-Daniels}, N. 2020, \mnras, 497, 1661

\bibitem[{{Bailes}(2022)}]{Bailes_FRB_review}
{Bailes}, M. 2022, Science, 378, abj3043

\bibitem[{{Barsdell} {et~al.}(2012){Barsdell}, {Bailes}, {Barnes}, \&
  {Fluke}}]{Heimdall}
{Barsdell}, B.~R., {Bailes}, M., {Barnes}, D.~G., \& {Fluke}, C.~J. 2012,
  \mnras, 422, 379

\bibitem[{{Barthelmy} {et~al.}(2008){Barthelmy}, {Baumgartner}, {Beardmore},
  {Burrows}, {Cummings}, {Evans}, {Gehrels}, {Godet}, {Guidorzi}, {Holland},
  {Kennea}, {Mangano}, {Mao}, {Marshall}, {O'Brien}, {Osborne}, {Pagani},
  {Page}, {Palmer}, {Perri}, {Sakamoto}, {Sbarufatti}, {Starling}, {Stratta},
  {Ukwatta}, \& {Vetere}}]{SGR0501_Barthelmy}
{Barthelmy}, S.~D., {Baumgartner}, W.~H., {Beardmore}, A.~P., {et~al.} 2008,
  The Astronomer's Telegram, 1676, 1

\bibitem[{{Bochenek} {et~al.}(2020){Bochenek}, {Ravi}, {Belov}, {Hallinan},
  {Kocz}, {Kulkarni}, \& {McKenna}}]{SGR1935_Bochenek}
{Bochenek}, C.~D., {Ravi}, V., {Belov}, K.~V., {et~al.} 2020, \nat, 587, 59

\bibitem[{{Borghese} {et~al.}(2022){Borghese}, {Coti Zelati}, {Israel},
  {Pilia}, {Burgay}, {Trudu}, {Zane}, {Turolla}, {Rea}, {Esposito},
  {Mereghetti}, {Tiengo}, \& {Possenti}}]{Borghese_1935}
{Borghese}, A., {Coti Zelati}, F., {Israel}, G.~L., {et~al.} 2022, \mnras, 516,
  602

\bibitem[{{Caleb} {et~al.}(2022){Caleb}, {Rajwade}, {Desvignes}, {Stappers},
  {Lyne}, {Weltevrede}, {Kramer}, {Levin}, \& {Surnis}}]{J1810_Giant_pulses}
{Caleb}, M., {Rajwade}, K., {Desvignes}, G., {et~al.} 2022, \mnras, 510, 1996

\bibitem[{{Camero} {et~al.}(2014){Camero}, {Papitto}, {Rea}, {Vigan{\`o}},
  {Pons}, {Tiengo}, {Mereghetti}, {Turolla}, {Esposito}, {Zane}, {Israel}, \&
  {G{\"o}tz}}]{SGR0501_Camero}
{Camero}, A., {Papitto}, A., {Rea}, N., {et~al.} 2014, \mnras, 438, 3291

\bibitem[{{Cameron} {et~al.}(2005){Cameron}, {Chandra}, {Ray}, {Kulkarni},
  {Frail}, {Wieringa}, {Nakar}, {Phinney}, {Miyazaki}, {Tsuboi}, {Okumura},
  {Kawai}, {Menten}, \& {Bertoldi}}]{Cameron_1806_prs}
{Cameron}, P.~B., {Chandra}, P., {Ray}, A., {et~al.} 2005, \nat, 434, 1112

\bibitem[{{Camilo} {et~al.}(2007){Camilo}, {Ransom}, {Halpern}, \&
  {Reynolds}}]{Camilo_1E1547_radio}
{Camilo}, F., {Ransom}, S.~M., {Halpern}, J.~P., \& {Reynolds}, J. 2007, \apjl,
  666, L93

\bibitem[{{Camilo} {et~al.}(2006){Camilo}, {Ransom}, {Halpern}, {Reynolds},
  {Helfand}, {Zimmerman}, \& {Sarkissian}}]{Camilo_XTEJ1810_radio}
{Camilo}, F., {Ransom}, S.~M., {Halpern}, J.~P., {et~al.} 2006, \nat, 442, 892

\bibitem[{{CHIME/FRB Collaboration} {et~al.}(2024){CHIME/FRB Collaboration},
  {Amiri}, {Andersen}, {Andrew}, {Bandura}, {Bhardwaj}, {Boyle}, {Brar},
  {Breitman}, {Cassanelli}, {Chawla}, {Cook}, {Curtin}, {Dobbs}, {Dong},
  {Eadie}, {Fonseca}, {Gaensler}, {Giri}, {Herrera-Martin}, {Hopkins}, {Ibik},
  {Joseph}, {Kaczmarek}, {Kader}, {Kaspi}, {Lanman}, {Lazda}, {Leung}, {Liu},
  {Masui}, {McKinven}, {Mena-Parra}, {Merryfield}, {Michilli}, {Ng}, {Nimmo},
  {Noble}, {Pandhi}, {Patel}, {Pearlman}, {Pen}, {Petroff}, {Pleunis},
  {Rafiei-Ravandi}, {Rahman}, {Ransom}, {Sand}, {Scholz}, {Shah}, {Shin},
  {Shpunarska}, {Siegel}, {Smith}, {Stairs}, {Stenning}, {Vanderlinde}, {Wang},
  {White}, \& {Wulf}}]{Baseband_chime_catalog}
{CHIME/FRB Collaboration}, {Amiri}, M., {Andersen}, B.~C., {et~al.} 2024, \apj,
  969, 145

\bibitem[{{CHIME/Frb Collaboration} {et~al.}(2023{\natexlab{a}}){CHIME/Frb
  Collaboration}, {Amiri}, {Andersen}, {Bandura}, {Berger}, {Bhardwaj},
  {Boyce}, {Boyle}, {Brar}, {Breitman}, {Cassanelli}, {Chawla}, {Chen},
  {Cliche}, {Cook}, {Cubranic}, {Curtin}, {Deng}, {Dobbs}, {Dong}, {Eadie},
  {Fandino}, {Fonseca}, {Gaensler}, {Giri}, {Good}, {Halpern}, {Hill},
  {Hinshaw}, {Josephy}, {Kaczmarek}, {Kader}, {Kania}, {Kaspi}, {Landecker},
  {Lang}, {Leung}, {Li}, {Lin}, {Masui}, {McKinven}, {Mena-Parra},
  {Merryfield}, {Meyers}, {Michilli}, {Milutinovic}, {Mirhosseini},
  {M{\"u}nchmeyer}, {Naidu}, {Newburgh}, {Ng}, {Patel}, {Pen}, {Petroff},
  {Pinsonneault-Marotte}, {Pleunis}, {Rafiei-Ravandi}, {Rahman}, {Ransom},
  {Renard}, {Sanghavi}, {Scholz}, {Shaw}, {Shin}, {Siegel}, {Sikora}, {Singh},
  {Smith}, {Stairs}, {Tan}, {Tendulkar}, {Vanderlinde}, {Wang}, {Wulf}, \&
  {Zwaniga}}]{CHIME_CAT1_corrected}
{CHIME/Frb Collaboration}, {Amiri}, M., {Andersen}, B.~C., {et~al.}
  2023{\natexlab{a}}, \apjs, 264, 53

\bibitem[{{CHIME/Frb Collaboration} {et~al.}(2023{\natexlab{b}}){CHIME/Frb
  Collaboration}, {Andersen}, {Bandura}, {Bhardwaj}, {Boyle}, {Brar},
  {Cassanelli}, {Chatterjee}, {Chawla}, {Cook}, {Curtin}, {Dobbs}, {Dong},
  {Faber}, {Fandino}, {Fonseca}, {Gaensler}, {Giri}, {Herrera-Martin}, {Hill},
  {Ibik}, {Josephy}, {Kaczmarek}, {Kader}, {Kaspi}, {Landecker}, {Lanman},
  {Lazda}, {Leung}, {Lin}, {Masui}, {McKinven}, {Mena-Parra}, {Meyers},
  {Michilli}, {Ng}, {Pandhi}, {Pearlman}, {Pen}, {Petroff}, {Pleunis},
  {Rafiei-Ravandi}, {Rahman}, {Ransom}, {Renard}, {Sand}, {Sanghavi}, {Scholz},
  {Shah}, {Shin}, {Siegel}, {Smith}, {Stairs}, {Su}, {Tendulkar},
  {Vanderlinde}, {Wang}, {Wulf}, \& {Zwaniga}}]{CHIME_repeaters_perc}
{CHIME/Frb Collaboration}, {Andersen}, B.~C., {Bandura}, K., {et~al.}
  2023{\natexlab{b}}, \apj, 947, 83

\bibitem[{{CHIME/FRB Collaboration} {et~al.}(2020){CHIME/FRB Collaboration},
  {Andersen}, {Bandura}, {Bhardwaj}, {Bij}, {Boyce}, {Boyle}, {Brar},
  {Cassanelli}, {Chawla}, {Chen}, {Cliche}, {Cook}, {Cubranic}, {Curtin},
  {Denman}, {Dobbs}, {Dong}, {Fandino}, {Fonseca}, {Gaensler}, {Giri}, {Good},
  {Halpern}, {Hill}, {Hinshaw}, {H{\"o}fer}, {Josephy}, {Kania}, {Kaspi},
  {Landecker}, {Leung}, {Li}, {Lin}, {Masui}, {McKinven}, {Mena-Parra},
  {Merryfield}, {Meyers}, {Michilli}, {Milutinovic}, {Mirhosseini},
  {M{\"u}nchmeyer}, {Naidu}, {Newburgh}, {Ng}, {Patel}, {Pen},
  {Pinsonneault-Marotte}, {Pleunis}, {Quine}, {Rafiei-Ravandi}, {Rahman},
  {Ransom}, {Renard}, {Sanghavi}, {Scholz}, {Shaw}, {Shin}, {Siegel}, {Singh},
  {Smegal}, {Smith}, {Stairs}, {Tan}, {Tendulkar}, {Tretyakov}, {Vanderlinde},
  {Wang}, {Wulf}, \& {Zwaniga}}]{SGR1935_CHIME}
{CHIME/FRB Collaboration}, {Andersen}, B.~C., {Bandura}, K.~M., {et~al.} 2020,
  \nat, 587, 54

\bibitem[{{Chrimes} {et~al.}(2025){Chrimes}, {Levan}, {Lyman}, {Borghese},
  {Dhillon}, {Esposito}, {Fraser}, {Fruchter}, {Gotz}, {Hounsell}, {Israel},
  {Kouveliotou}, {Mereghetti}, {Mignani}, {Perna}, {Rea}, {Skillen}, {Steeghs},
  {Tanvir}, {Wiersema}, {Wright}, \& {Zane}}]{0501_counter_Chrimes}
{Chrimes}, A.~A., {Levan}, A.~J., {Lyman}, J.~D., {et~al.} 2025, arXiv
  e-prints, arXiv:2504.08892

\bibitem[{{Cordes} \& {Lazio}(2002)}]{Pyne2001}
{Cordes}, J.~M. \& {Lazio}, T.~J.~W. 2002, arXiv e-prints, astro

\bibitem[{{Coti Zelati} {et~al.}(2018){Coti Zelati}, {Rea}, {Pons}, {Campana},
  \& {Esposito}}]{Coti_Zelati_outbursts}
{Coti Zelati}, F., {Rea}, N., {Pons}, J.~A., {Campana}, S., \& {Esposito}, P.
  2018, \mnras, 474, 961

\bibitem[{{Dall'Osso} \& {Stella}(2022)}]{Dallosso_magnetars}
{Dall'Osso}, S. \& {Stella}, L. 2022, in Astrophysics and Space Science
  Library, Vol. 465, Astrophysics and Space Science Library, ed.
  S.~{Bhattacharyya}, A.~{Papitto}, \& D.~{Bhattacharya}, 245--280

\bibitem[{{Davies} {et~al.}(2009){Davies}, {Figer}, {Kudritzki}, {Trombley},
  {Kouveliotou}, \& {Wachter}}]{SGR1900+14_DAVIES}
{Davies}, B., {Figer}, D.~F., {Kudritzki}, R.-P., {et~al.} 2009, \apj, 707, 844

\bibitem[{{Dhillon} {et~al.}(2011){Dhillon}, {Marsh}, {Littlefair},
  {Copperwheat}, {Hickman}, {Kerry}, {Levan}, {Rea}, {Savoury}, {Tanvir},
  {Turolla}, \& {Wiersema}}]{SGR0501_Dhillon}
{Dhillon}, V.~S., {Marsh}, T.~R., {Littlefair}, S.~P., {et~al.} 2011, \mnras,
  416, L16

\bibitem[{{Dib} \& {Kaspi}(2014)}]{1E_2259_Dib_2014}
{Dib}, R. \& {Kaspi}, V.~M. 2014, \apj, 784, 37

\bibitem[{{Durant} \& {van Kerkwijk}(2006)}]{4U0142+61_DURANT}
{Durant}, M. \& {van Kerkwijk}, M.~H. 2006, \apj, 650, 1070

\bibitem[{{Esposito} {et~al.}(2010){Esposito}, {Israel}, {Turolla}, {Tiengo},
  {G{\"o}tz}, {de Luca}, {Mignani}, {Zane}, {Rea}, {Testa}, {Caraveo}, {Chaty},
  {Mattana}, {Mereghetti}, {Pellizzoni}, \& {Romano}}]{Esposito_0418}
{Esposito}, P., {Israel}, G.~L., {Turolla}, R., {et~al.} 2010, \mnras, 405,
  1787

\bibitem[{{Esposito} {et~al.}(2008){Esposito}, {Israel}, {Zane}, {Senziani},
  {Starling}, {Rea}, {Palmer}, {Gehrels}, {Tiengo}, {de Luca}, {G{\"o}tz},
  {Mereghetti}, {Romano}, {Sakamoto}, {Barthelmy}, {Stella}, {Turolla},
  {Feroci}, \& {Mangano}}]{Esposito_2008_1627_outburst}
{Esposito}, P., {Israel}, G.~L., {Zane}, S., {et~al.} 2008, \mnras, 390, L34

\bibitem[{{Esposito} {et~al.}(2020){Esposito}, {Rea}, {Borghese}, {Coti
  Zelati}, {Vigan{\`o}}, {Israel}, {Tiengo}, {Ridolfi}, {Possenti}, {Burgay},
  {G{\"o}tz}, {Pintore}, {Stella}, {Dehman}, {Ronchi}, {Campana},
  {Garcia-Garcia}, {Graber}, {Mereghetti}, {Perna}, {Rodr{\'\i}guez Castillo},
  {Turolla}, \& {Zane}}]{Esposito_J1818_radio}
{Esposito}, P., {Rea}, N., {Borghese}, A., {et~al.} 2020, \apjl, 896, L30

\bibitem[{{Esposito} {et~al.}(2021){Esposito}, {Rea}, \&
  {Israel}}]{Esposito_magnetars_2021}
{Esposito}, P., {Rea}, N., \& {Israel}, G.~L. 2021, in Astrophysics and Space
  Science Library, Vol. 461, Timing Neutron Stars: Pulsations, Oscillations and
  Explosions, ed. T.~M. {Belloni}, M.~{M{\'e}ndez}, \& C.~{Zhang}, 97--142

\bibitem[{{Fahlman} \& {Gregory}(1981)}]{1E2259_Fahlman}
{Fahlman}, G.~G. \& {Gregory}, P.~C. 1981, \nat, 293, 202

\bibitem[{{Foley} {et~al.}(2012){Foley}, {Kouveliotou}, {Kaneko}, \&
  {Collazzi}}]{1E2259_Foley_outburst}
{Foley}, S., {Kouveliotou}, C., {Kaneko}, Y., \& {Collazzi}, A. 2012, GRB
  Coordinates Network, 13280, 1

\bibitem[{{Frail} {et~al.}(1999){Frail}, {Kulkarni}, \&
  {Bloom}}]{SGR1900_Frail}
{Frail}, D.~A., {Kulkarni}, S.~R., \& {Bloom}, J.~S. 1999, \nat, 398, 127

\bibitem[{{Gehrels}(1986)}]{Gehrels_1986}
{Gehrels}, N. 1986, \apj, 303, 336

\bibitem[{{G{\"o}{\v{g}}{\"u}{\c{s}}}
  {et~al.}(2001){G{\"o}{\v{g}}{\"u}{\c{s}}}, {Kouveliotou}, {Woods},
  {Thompson}, {Duncan}, \& {Briggs}}]{Gogus_burts_2001}
{G{\"o}{\v{g}}{\"u}{\c{s}}}, E., {Kouveliotou}, C., {Woods}, P.~M., {et~al.}
  2001, \apj, 558, 228

\bibitem[{{Gull{\'o}n} {et~al.}(2015){Gull{\'o}n}, {Pons}, {Miralles},
  {Vigan{\`o}}, {Rea}, \& {Perna}}]{Mag_pop_Gullon}
{Gull{\'o}n}, M., {Pons}, J.~A., {Miralles}, J.~A., {et~al.} 2015, \mnras, 454,
  615

\bibitem[{{Gupta} {et~al.}(2025){Gupta}, {Beniamini}, {Kumar}, \&
  {Finkelstein}}]{FRB_cosmo}
{Gupta}, O., {Beniamini}, P., {Kumar}, P., \& {Finkelstein}, S.~L. 2025, arXiv
  e-prints, arXiv:2501.09810

\bibitem[{{He} {et~al.}(2013){He}, {Ng}, \& {Kaspi}}]{HE_NH_DM_2013}
{He}, C., {Ng}, C.~Y., \& {Kaspi}, V.~M. 2013, \apj, 768, 64

\bibitem[{{Hulleman} {et~al.}(2000){Hulleman}, {van Kerkwijk}, \&
  {Kulkarni}}]{4U0142_Hulleman_2000}
{Hulleman}, F., {van Kerkwijk}, M.~H., \& {Kulkarni}, S.~R. 2000, \nat, 408,
  689

\bibitem[{{Hulleman} {et~al.}(2004){Hulleman}, {van Kerkwijk}, \&
  {Kulkarni}}]{4U0142_Hulleman_2004}
{Hulleman}, F., {van Kerkwijk}, M.~H., \& {Kulkarni}, S.~R. 2004, \aap, 416,
  1037

\bibitem[{{Hurley} {et~al.}(1999{\natexlab{a}}){Hurley}, {Cline}, {Mazets},
  {Barthelmy}, {Butterworth}, {Marshall}, {Palmer}, {Aptekar}, {Golenetskii},
  {Il'Inskii}, {Frederiks}, {McTiernan}, {Gold}, \&
  {Trombka}}]{SGR1900_Hurley_A}
{Hurley}, K., {Cline}, T., {Mazets}, E., {et~al.} 1999{\natexlab{a}}, \nat,
  397, 41

\bibitem[{{Hurley} {et~al.}(1999{\natexlab{b}}){Hurley}, {Li}, {Kouveliotou},
  {Murakami}, {Ando}, {Strohmayer}, {van Paradijs}, {Vrba}, {Luginbuhl},
  {Yoshida}, \& {Smith}}]{SGR1900_Hurley}
{Hurley}, K., {Li}, P., {Kouveliotou}, C., {et~al.} 1999{\natexlab{b}}, \apjl,
  510, L111

\bibitem[{{Ibrahim} {et~al.}(2024){Ibrahim}, {Borghese}, {Coti Zelati},
  {Parent}, {Marino}, {Ould-Boukattine}, {Rea}, {Ascenzi}, {Pacholski},
  {Mereghetti}, {Israel}, {Tiengo}, {Possenti}, {Burgay}, {Turolla}, {Zane},
  {Esposito}, {G{\"o}tz}, {Campana}, {Kirsten}, {Gawro{\'n}ski}, \&
  {Hessels}}]{Ibrahim_1935}
{Ibrahim}, A.~Y., {Borghese}, A., {Coti Zelati}, F., {et~al.} 2024, \apj, 965,
  87

\bibitem[{{Israel} {et~al.}(2021){Israel}, {Burgay}, {Rea}, {Esposito},
  {Possenti}, {Dall'Osso}, {Stella}, {Pilia}, {Tiengo}, {Ridnaia}, {Lien},
  {Frederiks}, \& {Bernardini}}]{Israel_1547_2021}
{Israel}, G.~L., {Burgay}, M., {Rea}, N., {et~al.} 2021, \apj, 907, 7

\bibitem[{{Israel} {et~al.}(2016){Israel}, {Esposito}, {Rea}, {Coti Zelati},
  {Tiengo}, {Campana}, {Mereghetti}, {Rodriguez Castillo}, {G{\"o}tz},
  {Burgay}, {Possenti}, {Zane}, {Turolla}, {Perna}, {Cannizzaro}, \&
  {Pons}}]{SGR1935+2154_ISRAEL}
{Israel}, G.~L., {Esposito}, P., {Rea}, N., {et~al.} 2016, \mnras, 457, 3448

\bibitem[{{Israel} {et~al.}(1994){Israel}, {Mereghetti}, \&
  {Stella}}]{4U0142_ISRAEL}
{Israel}, G.~L., {Mereghetti}, S., \& {Stella}, L. 1994, \apjl, 433, L25

\bibitem[{{Israel} {et~al.}(2008){Israel}, {Romano}, {Mangano}, {Dall'Osso},
  {Chincarini}, {Stella}, {Campana}, {Belloni}, {Tagliaferri}, {Blustin},
  {Sakamoto}, {Hurley}, {Zane}, {Moretti}, {Palmer}, {Guidorzi}, {Burrows},
  {Gehrels}, \& {Krimm}}]{Israel_bursts_2008}
{Israel}, G.~L., {Romano}, P., {Mangano}, V., {et~al.} 2008, \apj, 685, 1114

\bibitem[{{James} {et~al.}(2022){James}, {Prochaska}, {Macquart},
  {North-Hickey}, {Bannister}, \& {Dunning}}]{James_FRBpop}
{James}, C.~W., {Prochaska}, J.~X., {Macquart}, J.~P., {et~al.} 2022, \mnras,
  510, L18

\bibitem[{{Karuppusamy} {et~al.}(2010){Karuppusamy}, {Stappers}, \& {van
  Straten}}]{Giant_pulse_karu}
{Karuppusamy}, R., {Stappers}, B.~W., \& {van Straten}, W. 2010, \aap, 515, A36

\bibitem[{{Kaspi} \& {Beloborodov}(2017)}]{Kaspi_magnetars}
{Kaspi}, V.~M. \& {Beloborodov}, A.~M. 2017, \araa, 55, 261

\bibitem[{{Keane}(2018)}]{RRATS_pop}
{Keane}, E.~F. 2018, Nature Astronomy, 2, 865

\bibitem[{{Kirsten} {et~al.}(2021){Kirsten}, {Snelders}, {Jenkins}, {Nimmo},
  {van den Eijnden}, {Hessels}, {Gawro{\'n}ski}, \& {Yang}}]{SGR1935_Kirsten}
{Kirsten}, F., {Snelders}, M.~P., {Jenkins}, M., {et~al.} 2021, Nature
  Astronomy, 5, 414

\bibitem[{{Kothes} \& {Foster}(2012)}]{1E2259+586_KOTHES}
{Kothes}, R. \& {Foster}, T. 2012, \apjl, 746, L4

\bibitem[{{Kothes} {et~al.}(2018){Kothes}, {Sun}, {Gaensler}, \&
  {Reich}}]{SGR1935_Kothes_dist}
{Kothes}, R., {Sun}, X., {Gaensler}, B., \& {Reich}, W. 2018, \apj, 852, 54

\bibitem[{{Kouveliotou} {et~al.}(1999){Kouveliotou}, {Strohmayer}, {Hurley},
  {van Paradijs}, {Finger}, {Dieters}, {Woods}, {Thompson}, \&
  {Duncan}}]{SGR1900_Kouveliotou_1999}
{Kouveliotou}, C., {Strohmayer}, T., {Hurley}, K., {et~al.} 1999, \apjl, 510,
  L115

\bibitem[{{Kramer} {et~al.}(2007){Kramer}, {Stappers}, {Jessner}, {Lyne}, \&
  {Jordan}}]{Kramer_magnetar_emission}
{Kramer}, M., {Stappers}, B.~W., {Jessner}, A., {Lyne}, A.~G., \& {Jordan},
  C.~A. 2007, \mnras, 377, 107

\bibitem[{{Levin} {et~al.}(2010){Levin}, {Bailes}, {Bates}, {Bhat}, {Burgay},
  {Burke-Spolaor}, {D'Amico}, {Johnston}, {Keith}, {Kramer}, {Milia},
  {Possenti}, {Rea}, {Stappers}, \& {van Straten}}]{Levin_J1622_radio}
{Levin}, L., {Bailes}, M., {Bates}, S., {et~al.} 2010, \apjl, 721, L33

\bibitem[{{Lin} {et~al.}(2011){Lin}, {Kouveliotou}, {Baring}, {van der Horst},
  {Guiriec}, {Woods}, {G{\"o}{\v{g}}{\"u}{\c{s}}}, {Kaneko}, {Scargle},
  {Granot}, {Preece}, {von Kienlin}, {Chaplin}, {Watts}, {Wijers}, {Zhang},
  {Bhat}, {Finger}, {Gehrels}, {Harding}, {Kaper}, {Kaspi}, {Mcenery},
  {Meegan}, {Paciesas}, {Pe'er}, {Ramirez-Ruiz}, {van der Klis}, {Wachter}, \&
  {Wilson-Hodge}}]{SGR0501_Lin}
{Lin}, L., {Kouveliotou}, C., {Baring}, M.~G., {et~al.} 2011, \apj, 739, 87

\bibitem[{{Locatelli} {et~al.}(2020){Locatelli}, {Bernardi}, {Bianchi},
  {Chiello}, {Magro}, {Naldi}, {Pilia}, {Pupillo}, {Ridolfi}, {Setti}, \&
  {Vazza}}]{NC_Project_1}
{Locatelli}, N.~T., {Bernardi}, G., {Bianchi}, G., {et~al.} 2020, \mnras, 494,
  1229

\bibitem[{{Lorimer}(2011)}]{SIGPROC_Lorimer}
{Lorimer}, D.~R. 2011, {SIGPROC: Pulsar Signal Processing Programs},
  Astrophysics Source Code Library, record ascl:1107.016

\bibitem[{{Lorimer} {et~al.}(2007){Lorimer}, {Bailes}, {McLaughlin},
  {Narkevic}, \& {Crawford}}]{Lorimer2007}
{Lorimer}, D.~R., {Bailes}, M., {McLaughlin}, M.~A., {Narkevic}, D.~J., \&
  {Crawford}, F. 2007, Science, 318, 777

\bibitem[{{Lorimer} \& {Kramer}(2012)}]{Lorimer_and_Kramer}
{Lorimer}, D.~R. \& {Kramer}, M. 2012, {Handbook of Pulsar Astronomy}

\bibitem[{{Lundgren} {et~al.}(1995){Lundgren}, {Cordes}, {Ulmer}, {Matz},
  {Lomatch}, {Foster}, \& {Hankins}}]{Crab_giant_pulses}
{Lundgren}, S.~C., {Cordes}, J.~M., {Ulmer}, M., {et~al.} 1995, \apj, 453, 433

\bibitem[{{Margalit} {et~al.}(2020){Margalit}, {Beniamini}, {Sridhar}, \&
  {Metzger}}]{Margalit_2020}
{Margalit}, B., {Beniamini}, P., {Sridhar}, N., \& {Metzger}, B.~D. 2020,
  \apjl, 899, L27

\bibitem[{{Mazets} {et~al.}(1979{\natexlab{a}}){Mazets}, {Golenetskij}, \&
  {Guryan}}]{SGR1900_Mazets}
{Mazets}, E.~P., {Golenetskij}, S.~V., \& {Guryan}, Y.~A. 1979{\natexlab{a}},
  Soviet Astronomy Letters, 5, 343

\bibitem[{{Mazets} {et~al.}(1979{\natexlab{b}}){Mazets}, {Golentskii},
  {Ilinskii}, {Aptekar}, \& {Guryan}}]{Mazets_1979_Giant_flare}
{Mazets}, E.~P., {Golentskii}, S.~V., {Ilinskii}, V.~N., {Aptekar}, R.~L., \&
  {Guryan}, I.~A. 1979{\natexlab{b}}, \nat, 282, 587

\bibitem[{{Mereghetti}(2008)}]{Mereghetti_magnetars_2008}
{Mereghetti}, S. 2008, \aapr, 15, 225

\bibitem[{{Mereghetti} {et~al.}(2006){Mereghetti}, {Esposito}, {Tiengo},
  {Zane}, {Turolla}, {Stella}, {Israel}, {G{\"o}tz}, \&
  {Feroci}}]{SGR1900+14_MEREGHETTI}
{Mereghetti}, S., {Esposito}, P., {Tiengo}, A., {et~al.} 2006, \apj, 653, 1423

\bibitem[{{Mereghetti} {et~al.}(2020){Mereghetti}, {Savchenko}, {Ferrigno},
  {G{\"o}tz}, {Rigoselli}, {Tiengo}, {Bazzano}, {Bozzo}, {Coleiro},
  {Courvoisier}, {Doyle}, {Goldwurm}, {Hanlon}, {Jourdain}, {von Kienlin},
  {Lutovinov}, {Martin-Carrillo}, {Molkov}, {Natalucci}, {Onori}, {Panessa},
  {Rodi}, {Rodriguez}, {S{\'a}nchez-Fern{\'a}ndez}, {Sunyaev}, \&
  {Ubertini}}]{SGR1935_Mereghetti}
{Mereghetti}, S., {Savchenko}, V., {Ferrigno}, C., {et~al.} 2020, \apjl, 898,
  L29

\bibitem[{{Morello} {et~al.}(2022){Morello}, {Rajwade}, \& {Stappers}}]{IQRM}
{Morello}, V., {Rajwade}, K.~M., \& {Stappers}, B.~W. 2022, \mnras, 510, 1393

\bibitem[{{Morii} {et~al.}(2005){Morii}, {Kawai}, \&
  {Shibazaki}}]{4U0142_MORII}
{Morii}, M., {Kawai}, N., \& {Shibazaki}, N. 2005, \apj, 622, 544

\bibitem[{{Muno} {et~al.}(2008){Muno}, {Gaensler}, {Nechita}, {Miller}, \&
  {Slane}}]{Mag_pop_Muno}
{Muno}, M.~P., {Gaensler}, B.~M., {Nechita}, A., {Miller}, J.~M., \& {Slane},
  P.~O. 2008, \apj, 680, 639

\bibitem[{{Nimmo} {et~al.}(2022){Nimmo}, {Hessels}, {Kirsten}, {Keimpema},
  {Cordes}, {Snelders}, {Hewitt}, {Karuppusamy}, {Archibald}, {Bezrukovs},
  {Bhardwaj}, {Blaauw}, {Buttaccio}, {Cassanelli}, {Conway}, {Corongiu},
  {Feiler}, {Fonseca}, {Forss{\'e}n}, {Gawro{\'n}ski}, {Giroletti}, {Kharinov},
  {Leung}, {Lindqvist}, {Maccaferri}, {Marcote}, {Masui}, {Mckinven},
  {Melnikov}, {Michilli}, {Mikhailov}, {Ng}, {Orbidans}, {Ould-Boukattine},
  {Paragi}, {Pearlman}, {Petroff}, {Rahman}, {Scholz}, {Shin}, {Smith},
  {Stairs}, {Surcis}, {Tendulkar}, {Vlemmings}, {Wang}, {Yang}, \&
  {Yuan}}]{Nimmo_2022_frb}
{Nimmo}, K., {Hessels}, J.~W.~T., {Kirsten}, F., {et~al.} 2022, Nature
  Astronomy, 6, 393

\bibitem[{{Olausen} \& {Kaspi}(2014)}]{Olausen_McGill_Cat}
{Olausen}, S.~A. \& {Kaspi}, V.~M. 2014, \apjs, 212, 6

\bibitem[{{Palmer} {et~al.}(2005){Palmer}, {Barthelmy}, {Gehrels}, {Kippen},
  {Cayton}, {Kouveliotou}, {Eichler}, {Wijers}, {Woods}, {Granot}, {Lyubarsky},
  {Ramirez-Ruiz}, {Barbier}, {Chester}, {Cummings}, {Fenimore}, {Finger},
  {Gaensler}, {Hullinger}, {Krimm}, {Markwardt}, {Nousek}, {Parsons}, {Patel},
  {Sakamoto}, {Sato}, {Suzuki}, \& {Tueller}}]{Palmer_MGF_SGR1806-20}
{Palmer}, D.~M., {Barthelmy}, S., {Gehrels}, N., {et~al.} 2005, \nat, 434, 1107

\bibitem[{{Pelliciari} {et~al.}(2024){Pelliciari}, {Bernardi}, {Pilia},
  {Naldi}, {Maccaferri}, {Verrecchia}, {Casentini}, {Perri}, {Kirsten},
  {Bianchi}, {Bortolotti}, {Bruno}, {Dallacasa}, {Esposito}, {Geminardi},
  {Giarratana}, {Giroletti}, {Lulli}, {Maccaferri}, {Magro}, {Mattana},
  {Perini}, {Pupillo}, {Roma}, {Schiaffino}, {Setti}, {Tavani}, {Trudu}, \&
  {Zanichelli}}]{NC_Project_4}
{Pelliciari}, D., {Bernardi}, G., {Pilia}, M., {et~al.} 2024, \aap, 690, A219

\bibitem[{{Pelliciari} {et~al.}(2023){Pelliciari}, {Bernardi}, {Pilia},
  {Naldi}, {Pupillo}, {Trudu}, {Addis}, {Bianchi}, {Bortolotti}, {Dallacasa},
  {Lulli}, {Maccaferri}, {Magro}, {Mattana}, {Perini}, {Roma}, {Schiaffino},
  {Setti}, {Tavani}, {Verrecchia}, \& {Casentini}}]{NC_Project_3}
{Pelliciari}, D., {Bernardi}, G., {Pilia}, M., {et~al.} 2023, \aap, 674, A223

\bibitem[{{Pizzocaro} {et~al.}(2019){Pizzocaro}, {Tiengo}, {Mereghetti},
  {Turolla}, {Esposito}, {Stella}, {Zane}, {Rea}, {Coti Zelati}, \&
  {Israel}}]{1E2259+586_PIZZOCARO}
{Pizzocaro}, D., {Tiengo}, A., {Mereghetti}, S., {et~al.} 2019, \aap, 626, A39

\bibitem[{{Platts} {et~al.}(2019){Platts}, {Weltman}, {Walters}, {Tendulkar},
  {Gordin}, \& {Kandhai}}]{Platts_FRB_models}
{Platts}, E., {Weltman}, A., {Walters}, A., {et~al.} 2019, \physrep, 821, 1

\bibitem[{{Pleunis} {et~al.}(2021){Pleunis}, {Good}, {Kaspi}, {Mckinven},
  {Ransom}, {Scholz}, {Bandura}, {Bhardwaj}, {Boyle}, {Brar}, {Cassanelli},
  {Chawla}, {(Adam) Dong}, {Fonseca}, {Gaensler}, {Josephy}, {Kaczmarek},
  {Leung}, {Lin}, {Masui}, {Mena-Parra}, {Michilli}, {Ng}, {Patel},
  {Rafiei-Ravandi}, {Rahman}, {Sanghavi}, {Shin}, {Smith}, {Stairs}, \&
  {Tendulkar}}]{Pleunis_2021_frbcat}
{Pleunis}, Z., {Good}, D.~C., {Kaspi}, V.~M., {et~al.} 2021, \apj, 923, 1

\bibitem[{{Popov} {et~al.}(2018){Popov}, {Postnov}, \&
  {Pshirkov}}]{Popov_magnetars_FRB}
{Popov}, S.~B., {Postnov}, K.~A., \& {Pshirkov}, M.~S. 2018, Physics Uspekhi,
  61, 965

\bibitem[{{Ranasinghe} {et~al.}(2018){Ranasinghe}, {Leahy}, \&
  {Tian}}]{SGR1935_Ranasinghe_dist}
{Ranasinghe}, S., {Leahy}, D.~A., \& {Tian}, W. 2018, Open Physics Journal, 4,
  1

\bibitem[{{Rea} \& {Esposito}(2011)}]{Rea_esposito_outbursts}
{Rea}, N. \& {Esposito}, P. 2011, in Astrophysics and Space Science
  Proceedings, Vol.~21, High-Energy Emission from Pulsars and their Systems,
  ed. D.~F. {Torres} \& N.~{Rea}, 247

\bibitem[{{Rea} {et~al.}(2013{\natexlab{a}}){Rea}, {Esposito}, {Pons},
  {Turolla}, {Torres}, {Israel}, {Possenti}, {Burgay}, {Vigan{\`o}}, {Papitto},
  {Perna}, {Stella}, {Ponti}, {Baganoff}, {Haggard}, {Camero-Arranz}, {Zane},
  {Minter}, {Mereghetti}, {Tiengo}, {Sch{\"o}del}, {Feroci}, {Mignani}, \&
  {G{\"o}tz}}]{Rea_J1745_radio}
{Rea}, N., {Esposito}, P., {Pons}, J.~A., {et~al.} 2013{\natexlab{a}}, \apjl,
  775, L34

\bibitem[{{Rea} {et~al.}(2010){Rea}, {Esposito}, {Turolla}, {Israel}, {Zane},
  {Stella}, {Mereghetti}, {Tiengo}, {G{\"o}tz}, {G{\"o}{\u{g}}{\"u}{\c{s}}}, \&
  {Kouveliotou}}]{SGR0418_REA_2010}
{Rea}, N., {Esposito}, P., {Turolla}, R., {et~al.} 2010, Science, 330, 944

\bibitem[{{Rea} {et~al.}(2013{\natexlab{b}}){Rea}, {Israel}, {Pons}, {Turolla},
  {Vigan{\`o}}, {Zane}, {Esposito}, {Perna}, {Papitto}, {Terreran}, {Tiengo},
  {Salvetti}, {Girart}, {Palau}, {Possenti}, {Burgay},
  {G{\"o}{\u{g}}{\"u}{\c{s}}}, {Caliandro}, {Kouveliotou}, {G{\"o}tz},
  {Mignani}, {Ratti}, \& {Stella}}]{SGR0418_REA_2013}
{Rea}, N., {Israel}, G.~L., {Pons}, J.~A., {et~al.} 2013{\natexlab{b}}, \apj,
  770, 65

\bibitem[{{Rea} {et~al.}(2009){Rea}, {Israel}, {Turolla}, {Esposito},
  {Mereghetti}, {G{\"o}tz}, {Zane}, {Tiengo}, {Hurley}, {Feroci}, {Still},
  {Yershov}, {Winkler}, {Perna}, {Bernardini}, {Ubertini}, {Stella}, {Campana},
  {van der Klis}, \& {Woods}}]{SGR0501_Rea}
{Rea}, N., {Israel}, G.~L., {Turolla}, R., {et~al.} 2009, \mnras, 396, 2419

\bibitem[{{Rea} {et~al.}(2007){Rea}, {Nichelli}, {Israel}, {Perna},
  {Oosterbroek}, {Parmar}, {Turolla}, {Campana}, {Stella}, {Zane}, \&
  {Angelini}}]{4U0142_REA}
{Rea}, N., {Nichelli}, E., {Israel}, G.~L., {et~al.} 2007, \mnras, 381, 293

\bibitem[{{Rea} {et~al.}(2014){Rea}, {Vigan{\`o}}, {Israel}, {Pons}, \&
  {Torres}}]{3XMMJ18+00_REA}
{Rea}, N., {Vigan{\`o}}, D., {Israel}, G.~L., {Pons}, J.~A., \& {Torres}, D.~F.
  2014, \apjl, 781, L17

\bibitem[{{Rehan} \& {Ibrahim}(2025)}]{SGR1935_Rehan}
{Rehan}, N.~S. \& {Ibrahim}, A.~I. 2025, \apjs, 276, 60

\bibitem[{{Remazeilles} {et~al.}(2015){Remazeilles}, {Dickinson}, {Banday},
  {Bigot-Sazy}, \& {Ghosh}}]{Map_408MHz_2014}
{Remazeilles}, M., {Dickinson}, C., {Banday}, A.~J., {Bigot-Sazy}, M.~A., \&
  {Ghosh}, T. 2015, \mnras, 451, 4311

\bibitem[{{Ryder} {et~al.}(2023){Ryder}, {Bannister}, {Bhandari}, {Deller},
  {Ekers}, {Glowacki}, {Gordon}, {Gourdji}, {James}, {Kilpatrick}, {Lu},
  {Marnoch}, {Moss}, {Prochaska}, {Qiu}, {Sadler}, {Simha}, {Sammons}, {Scott},
  {Tejos}, \& {Shannon}}]{Ryder_2023_Z1}
{Ryder}, S.~D., {Bannister}, K.~W., {Bhandari}, S., {et~al.} 2023, Science,
  382, 294

\bibitem[{{Sakamoto} {et~al.}(2011{\natexlab{a}}){Sakamoto}, {Barbier},
  {Barthelmy}, {Cummings}, {Fenimore}, {Gehrels}, {Krimm}, {Markwardt},
  {Palmer}, {Parsons}, {Sato}, {Stamatikos}, \&
  {Tueller}}]{SGR2013+34_SAKAMOTO}
{Sakamoto}, T., {Barbier}, L., {Barthelmy}, S.~D., {et~al.} 2011{\natexlab{a}},
  Advances in Space Research, 47, 1346

\bibitem[{{Sakamoto} {et~al.}(2011{\natexlab{b}}){Sakamoto}, {Barthelmy},
  {Baumgartner}, {Cummings}, {Fenimore}, {Gehrels}, {Krimm}, {Markwardt},
  {Palmer}, {Parsons}, {Sato}, {Stamatikos}, {Tueller}, {Ukwatta}, \&
  {Zhang}}]{SGR2013_SWIFT_CAT}
{Sakamoto}, T., {Barthelmy}, S.~D., {Baumgartner}, W.~H., {et~al.}
  2011{\natexlab{b}}, \apjs, 195, 2

\bibitem[{{Tavani} {et~al.}(2021){Tavani}, {Casentini}, {Ursi}, {Verrecchia},
  {Addis}, {Antonelli}, {Argan}, {Barbiellini}, {Baroncelli}, {Bernardi},
  {Bianchi}, {Bulgarelli}, {Caraveo}, {Cardillo}, {Cattaneo}, {Chen}, {Costa},
  {Del Monte}, {Di Cocco}, {Di Persio}, {Donnarumma}, {Evangelista}, {Feroci},
  {Ferrari}, {Fioretti}, {Fuschino}, {Galli}, {Gianotti}, {Giuliani},
  {Labanti}, {Lazzarotto}, {Lipari}, {Longo}, {Lucarelli}, {Magro},
  {Marisaldi}, {Mereghetti}, {Morelli}, {Morselli}, {Naldi}, {Pacciani},
  {Parmiggiani}, {Paoletti}, {Pellizzoni}, {Perri}, {Perotti}, {Piano},
  {Picozza}, {Pilia}, {Pittori}, {Puccetti}, {Pupillo}, {Rapisarda},
  {Rappoldi}, {Rubini}, {Setti}, {Soffitta}, {Trifoglio}, {Trois},
  {Vercellone}, {Vittorini}, {Giommi}, \& {D'Amico}}]{1935_Tavani}
{Tavani}, M., {Casentini}, C., {Ursi}, A., {et~al.} 2021, Nature Astronomy, 5,
  401

\bibitem[{{Tendulkar} {et~al.}(2013){Tendulkar}, {Cameron}, \&
  {Kulkarni}}]{Tendulkar_2013_4U0142_1E2259}
{Tendulkar}, S.~P., {Cameron}, P.~B., \& {Kulkarni}, S.~R. 2013, \apj, 772, 31

\bibitem[{{Trudu} {et~al.}(2022){Trudu}, {Pilia}, {Bernardi}, {Addis},
  {Bianchi}, {Magro}, {Naldi}, {Pelliciari}, {Pupillo}, {Setti}, {Bortolotti},
  {Casentini}, {Dallacasa}, {Gajjar}, {Locatelli}, {Lulli}, {Maccaferri},
  {Mattana}, {Michilli}, {Perini}, {Possenti}, {Roma}, {Schiaffino}, {Tavani},
  \& {Verrecchia}}]{NC_Project_2}
{Trudu}, M., {Pilia}, M., {Bernardi}, G., {et~al.} 2022, \mnras, 513, 1858

\bibitem[{{Turolla} {et~al.}(2015){Turolla}, {Zane}, \&
  {Watts}}]{Turolla_magnetars_2015}
{Turolla}, R., {Zane}, S., \& {Watts}, A.~L. 2015, Reports on Progress in
  Physics, 78, 116901

\bibitem[{{van der Horst} {et~al.}(2010){van der Horst}, {Connaughton},
  {Kouveliotou}, {G{\"o}{\v{g}}{\"u}{\c{s}}}, {Kaneko}, {Wachter}, {Briggs},
  {Granot}, {Ramirez-Ruiz}, {Woods}, {Aptekar}, {Barthelmy}, {Cummings},
  {Finger}, {Frederiks}, {Gehrels}, {Gelino}, {Gelino}, {Golenetskii},
  {Hurley}, {Krimm}, {Mazets}, {McEnery}, {Meegan}, {Oleynik}, {Palmer},
  {Pal'shin}, {Pe'er}, {Svinkin}, {Ulanov}, {van der Klis}, {von Kienlin},
  {Watts}, \& {Wilson-Hodge}}]{SGR0418_VDH_2010}
{van der Horst}, A.~J., {Connaughton}, V., {Kouveliotou}, C., {et~al.} 2010,
  \apjl, 711, L1

\bibitem[{{Woods} {et~al.}(2004){Woods}, {Kaspi}, {Thompson}, {Gavriil},
  {Marshall}, {Chakrabarty}, {Flanagan}, {Heyl}, \&
  {Hernquist}}]{Woods_2004_1E2259_outburst}
{Woods}, P.~M., {Kaspi}, V.~M., {Thompson}, C., {et~al.} 2004, \apj, 605, 378

\bibitem[{{Xie} {et~al.}(2024){Xie}, {Han}, {Yang}, {Jing}, {Zhou}, {Su},
  {Yan}, {Wang}, {Cai}, {Wang}, \& {Wang}}]{FAST_1935_3XMM}
{Xie}, L., {Han}, J.~L., {Yang}, Z.~L., {et~al.} 2024, arXiv e-prints,
  arXiv:2411.15960

\bibitem[{{Xu} {et~al.}(2023){Xu}, {Feng}, {Li}, {Wang}, {Zhang}, {Xie},
  {Chen}, {Wang}, {Kang}, {Hu}, {Zheng}, {Tsai}, {Chen}, \&
  {Zhou}}]{Blinkverse_cat}
{Xu}, J., {Feng}, Y., {Li}, D., {et~al.} 2023, Universe, 9, 330

\bibitem[{{Xu} {et~al.}(2006){Xu}, {Reid}, {Zheng}, \& {Menten}}]{Perseus_Xu}
{Xu}, Y., {Reid}, M.~J., {Zheng}, X.~W., \& {Menten}, K.~M. 2006, Science, 311,
  54

\bibitem[{{Yao} {et~al.}(2017){Yao}, {Manchester}, \& {Wang}}]{YMW16}
{Yao}, J.~M., {Manchester}, R.~N., \& {Wang}, N. 2017, \apj, 835, 29

\bibitem[{Zhang(2023)}]{Zhang_FRB_review}
Zhang, B. 2023, Rev. Mod. Phys., 95, 035005

\bibitem[{{Zhong} {et~al.}(2020){Zhong}, {Dai}, {Zhang}, \&
  {Deng}}]{SGR1935_Zhong}
{Zhong}, S.-Q., {Dai}, Z.-G., {Zhang}, H.-M., \& {Deng}, C.-M. 2020, \apjl,
  898, L5

\bibitem[{{Zhou} {et~al.}(2014){Zhou}, {Chen}, {Li}, {Safi-Harb}, {Mendez},
  {Terada}, {Sun}, \& {Ge}}]{3XMMJ18_Zhou}
{Zhou}, P., {Chen}, Y., {Li}, X.-D., {et~al.} 2014, \apjl, 781, L16

\bibitem[{{Zhu} {et~al.}(2023){Zhu}, {Xu}, {Zhou}, {Lin}, {Wang}, {Wang},
  {Zhang}, {Niu}, {Chen}, {Li}, {Meng}, {Lee}, {Zhang}, {Feng}, {Ge},
  {G{\"o}{\u{g}}{\"u}{\c{s}}}, {Guan}, {Han}, {Jiang}, {Jiang}, {Kouveliotou},
  {Li}, {Miao}, {Miao}, {Men}, {Niu}, {Wang}, {Wang}, {Xu}, {Xu}, {Xue},
  {Yang}, {Yu}, {Yuan}, {Yue}, {Zhang}, \& {Zhang}}]{1935_pulsed_emission_FAST}
{Zhu}, W., {Xu}, H., {Zhou}, D., {et~al.} 2023, Science Advances, 9, eadf6198

\end{thebibliography}

\bibliographystyle{aa}

\begin{appendix}
\section{Individual magnetar results} \label{Appendix:single_plots}
\begin{figure*}[htbp]
    \centering
        \includegraphics[width=0.45\linewidth,height=0.28\linewidth]{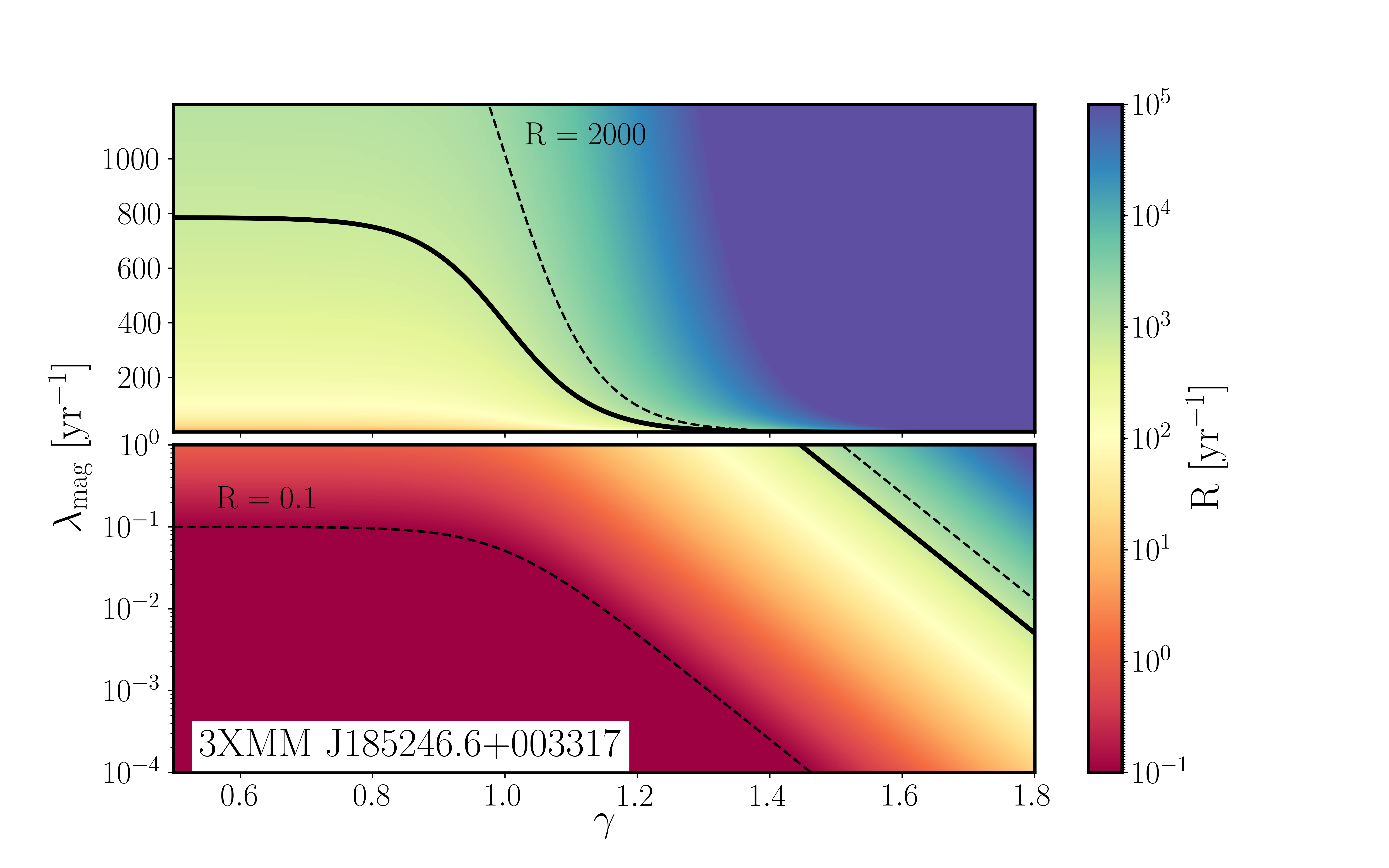}
        \includegraphics[width=0.45\linewidth,height=0.28\linewidth]{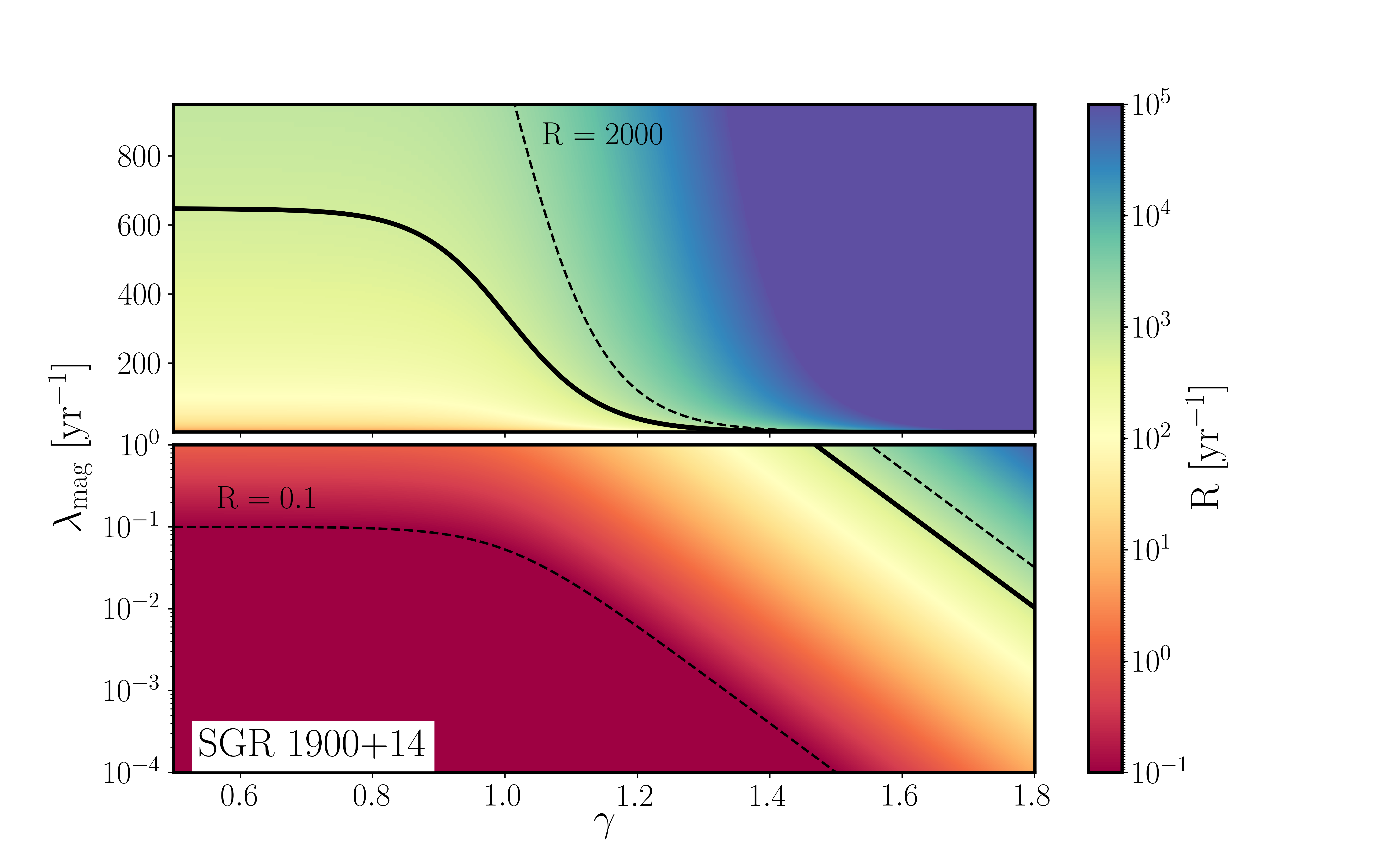}
        \includegraphics[width=0.45\linewidth,height=0.28\linewidth]{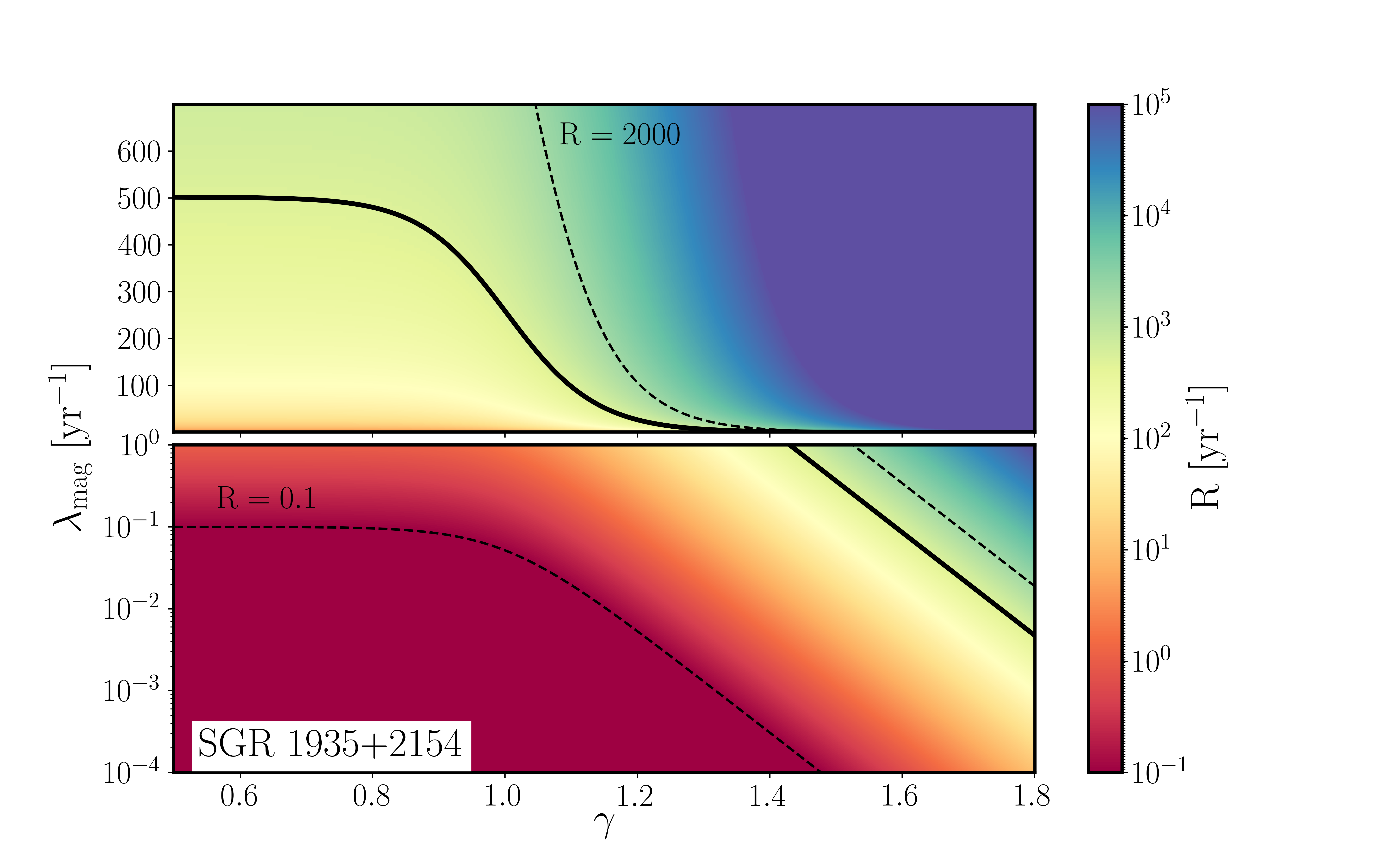}
        \includegraphics[width=0.45\linewidth,height=0.28\linewidth]{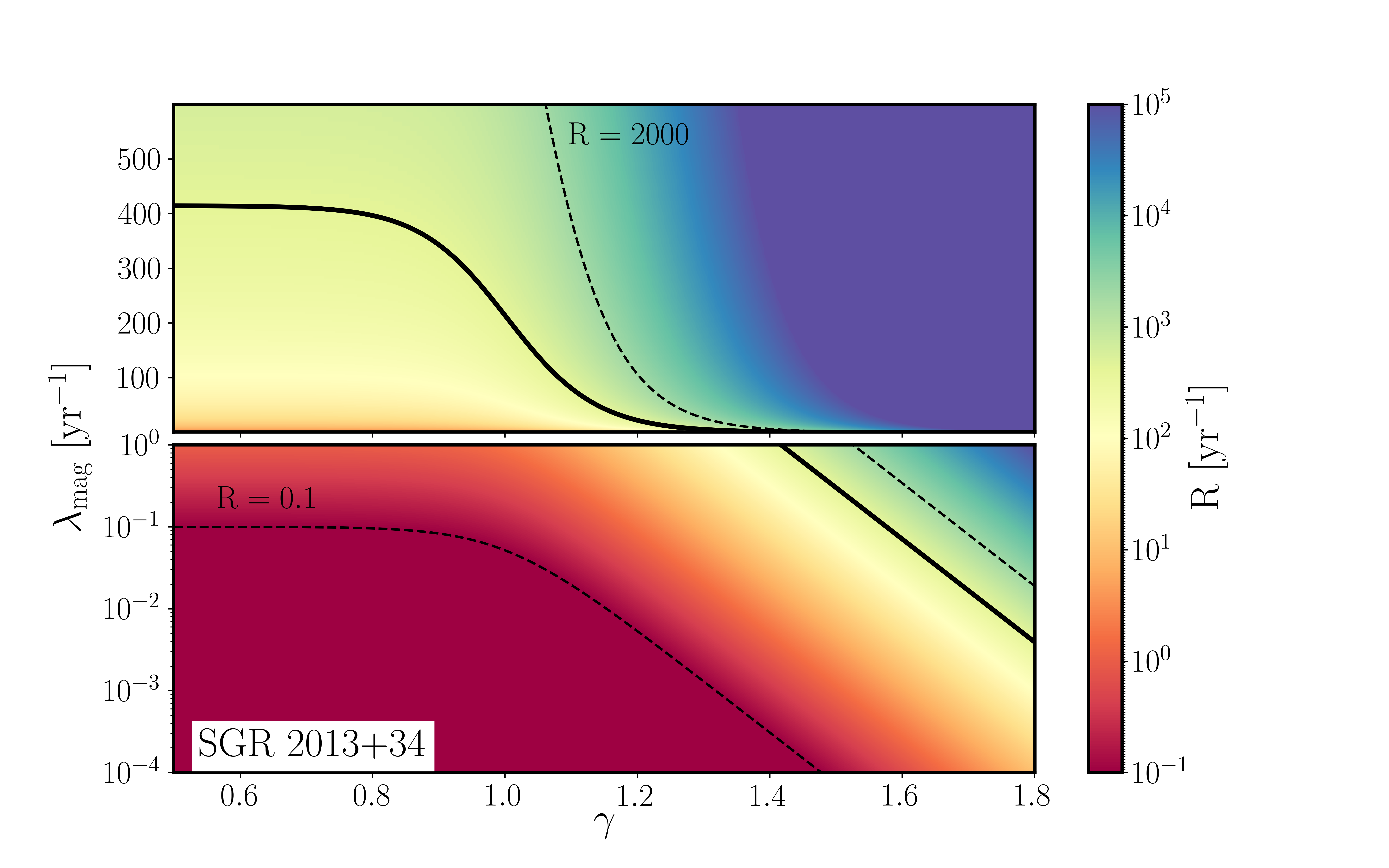}
        \includegraphics[width=0.45\linewidth,height=0.28\linewidth]{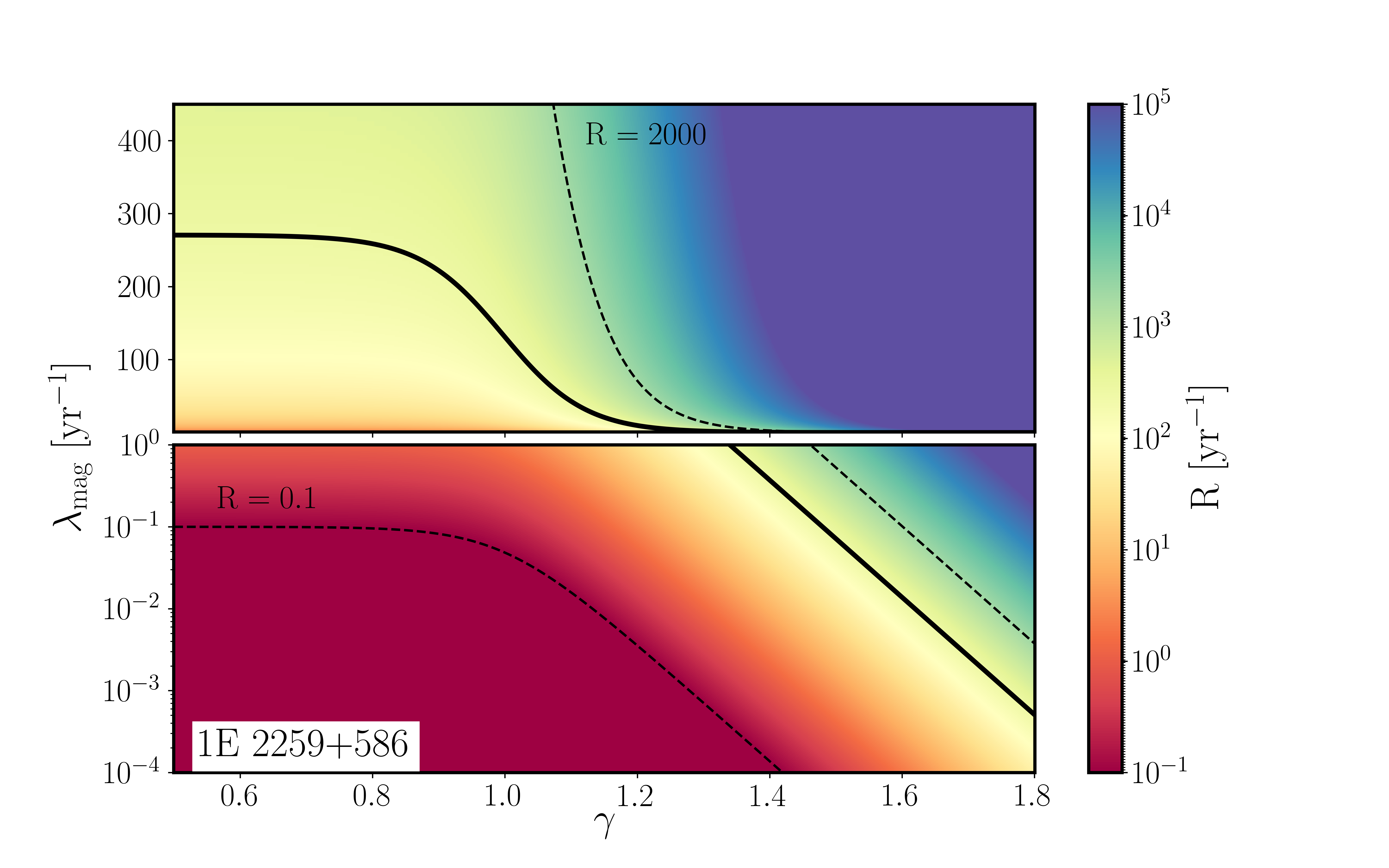}
        \includegraphics[width=0.45\linewidth,height=0.28\linewidth]{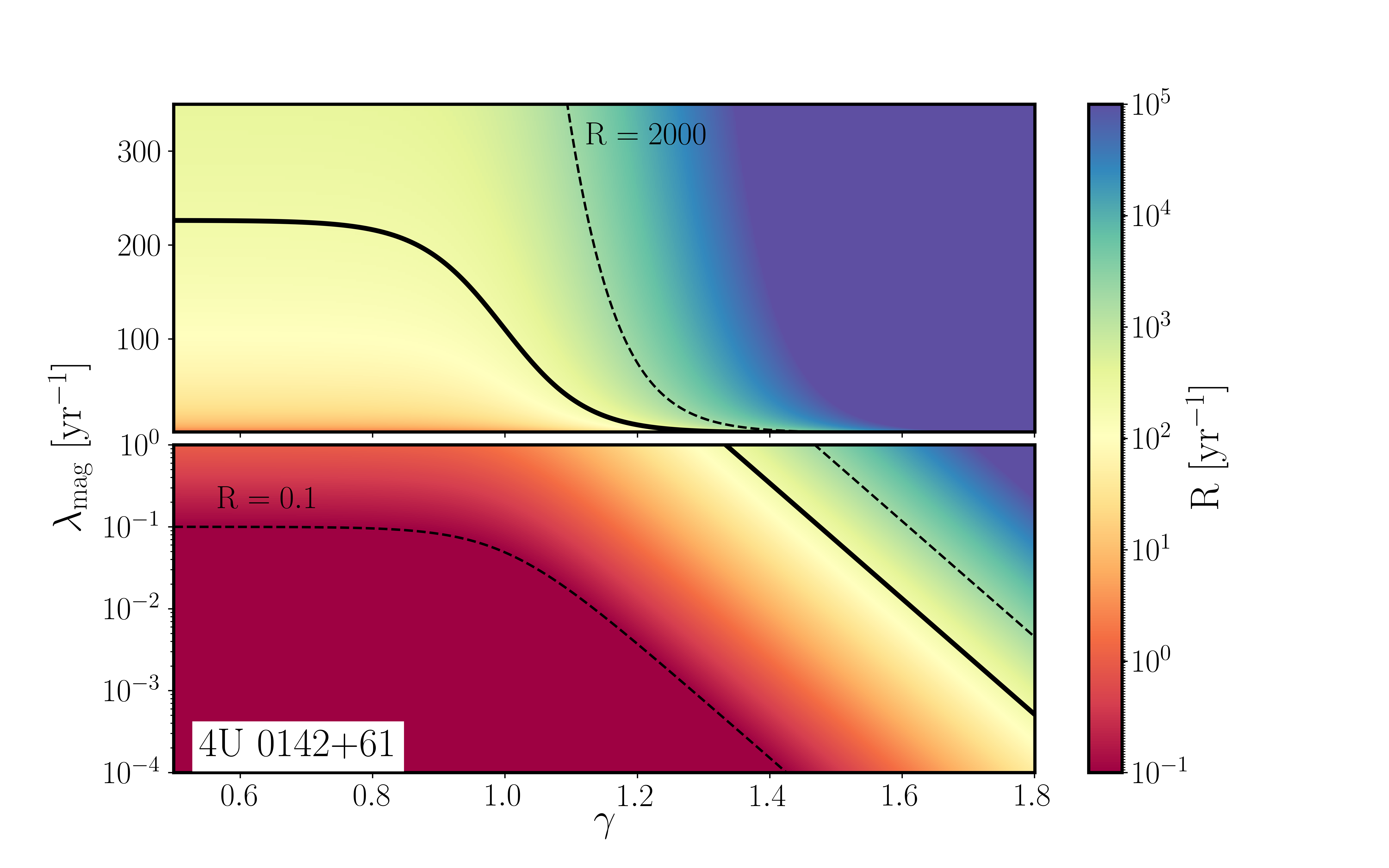}
        \includegraphics[width=0.45\linewidth,height=0.28\linewidth]{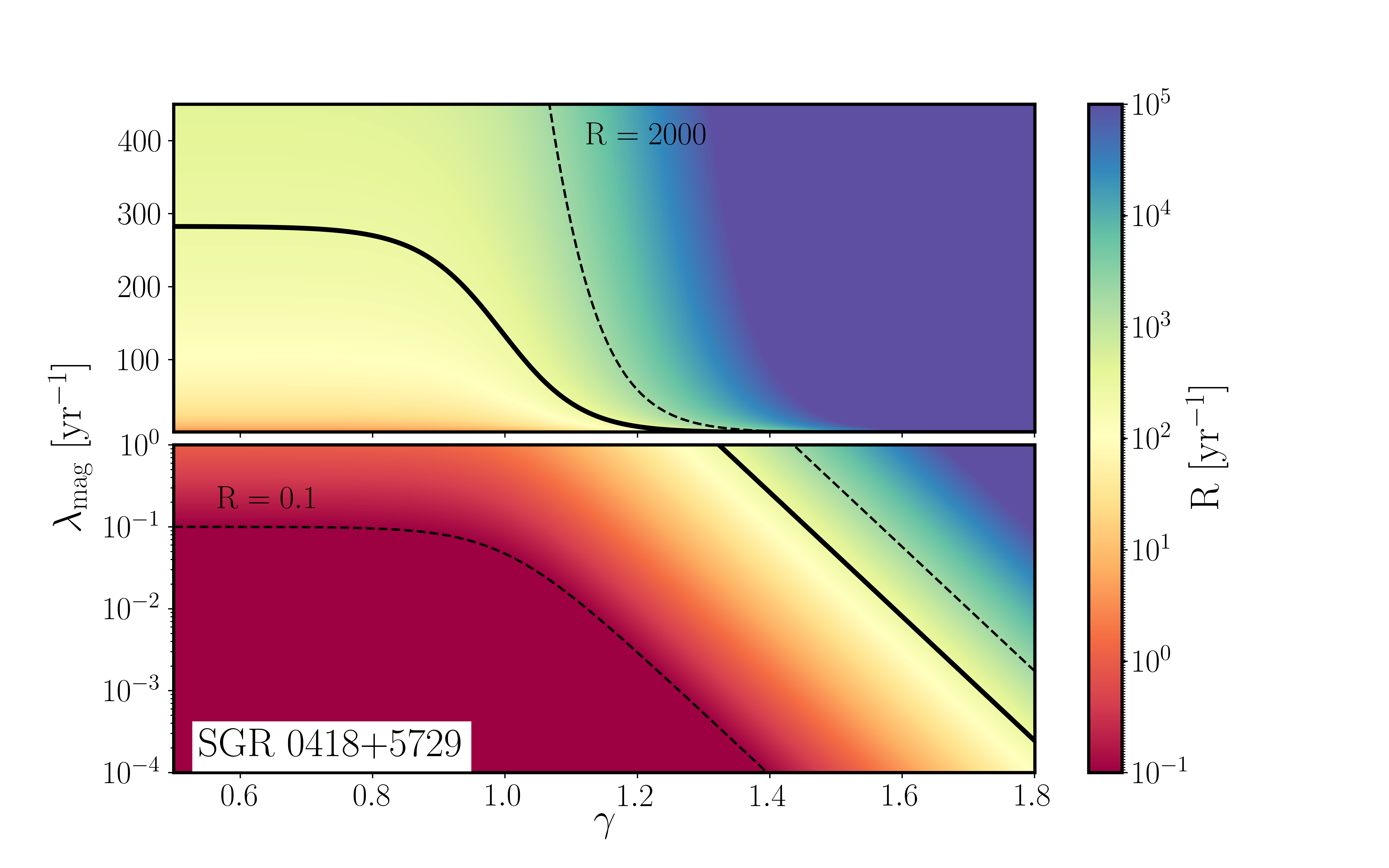}
        \includegraphics[width=0.45\linewidth,height=0.28\linewidth]{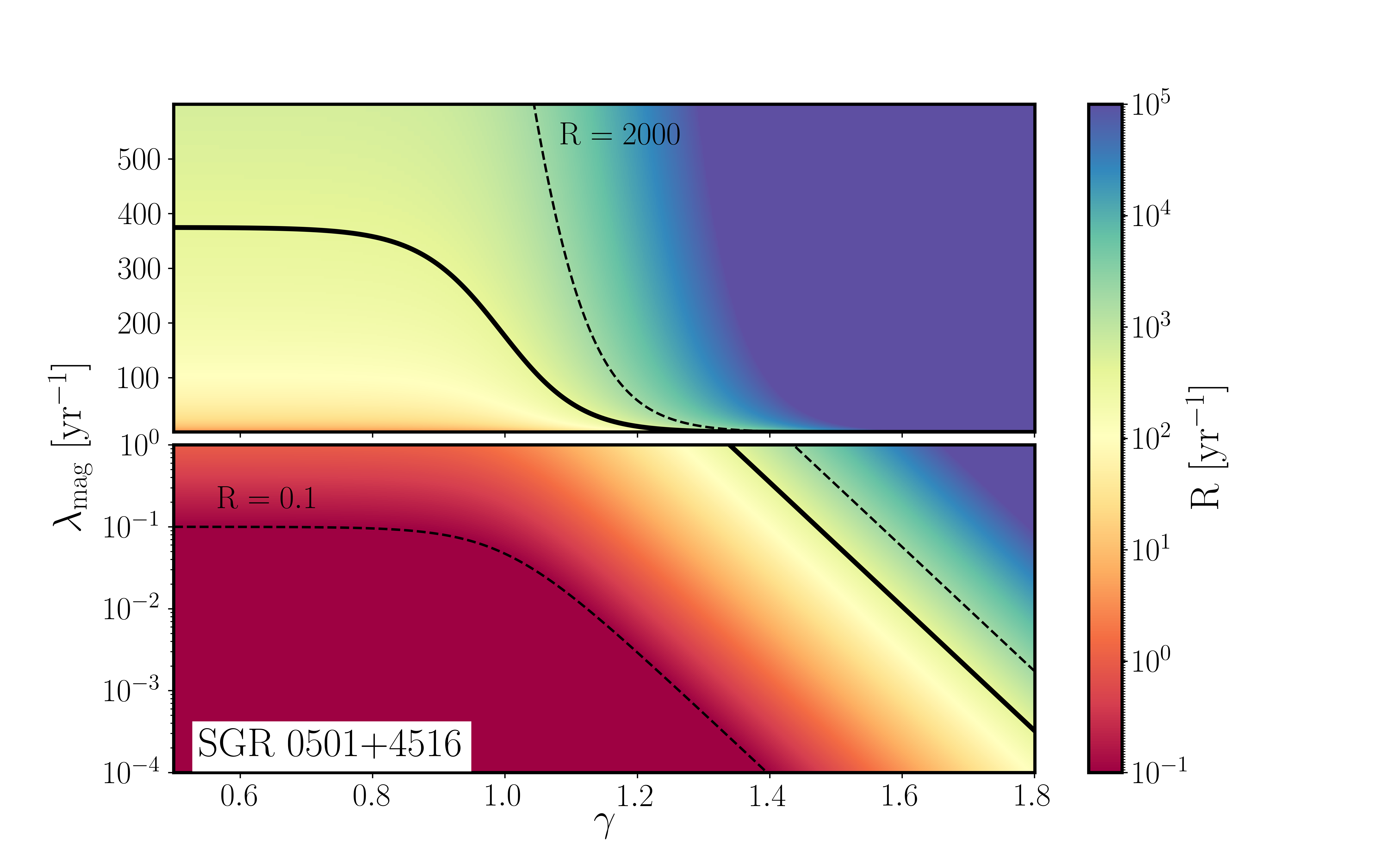}
        \caption{The plots show a graphical representation of our power-law energy distribution model for each magnetar in our sample. The x-axis represents the slope of the power-law energy distribution $\gamma$, while on the y-axis there is the number of events with energy greater than the energy emitted by the FRB-like event of SGR\,J1935+2154 $\lambda_{\rm{mag}}$.
        The rate of events obtained in every case, are showed using a fixed logarithmic color code presented in the colorbars.
        The black line represents the upper limits, obtained with our observing campaign, on the possible combinations of $\gamma$ and $\lambda_{\rm{mag}}$, while the black dashed lines show reference values for graphical purposes. The difference between upper and lower panels is the scale of the y-axis, the upper panels are in linear scale and the lower panels in logarithmic scale.}
        \label{fig:single_magnetars}
\end{figure*}

In Fig. \ref{fig:single_magnetars}, we report on our results for each individual target, obtained using Eq. \ref{EQ:E_rate_3}.
\end{appendix}

\end{document}